\documentclass[journal,letterpaper]{IEEEtran}

\pdfoutput=1
\usepackage{cite}
\usepackage{graphicx}
\usepackage{subfigure}
\usepackage{amssymb,amsmath} 
\usepackage{multirow}

\begin{document}

\title{Photon elastic scattering simulation: validation and improvements to Geant4 }

\author{Matej Bati\v{c}, Gabriela Hoff, Maria Grazia Pia and Paolo Saracco
\thanks{Manuscript received 28 February 2012.}
\thanks{This work has been partly funded by CNPq BEX6460/10-0 grant, Brazil.}
\thanks{M. Bati\v{c} is with INFN Sezione di Genova, Genova, Italy 
             (e-mail: Matej.Batic@ge.infn.it); he is on leave from 
             Jozef Stefan Institute, 1000 Ljubljana, Slovenia.}
\thanks{G. Hoff is with INFN Sezione di Genova, Genova, Italy 
             (e-mail: Gabriela.Hoff@ge.infn.it); she is on leave from 
             Pontificia Universidade Catolica do Rio Grande do Sul, Brazil.}
\thanks{M. G. Pia and Paolo Saracco are  with INFN Sezione di Genova, Via Dodecaneso 33, I-16146 Genova, Italy 
	(phone: +39 010 3536328, fax: +39 010 313358,
	MariaGrazia.Pia@ge.infn.it,Paolo.Saracco@ge.infn.it).}
}

\maketitle

\begin{abstract}
Several models for the simulation of photon elastic scattering are
quantitatively evaluated with respect to a large collection of experimental data 
retrieved from the literature.
They include models based on the form factor approximation, on S-matrix
calculations and on analytical parameterizations; they exploit publicly
available data libraries and tabulations of theoretical calculations.
Some of these models are currently implemented in general purpose Monte Carlo
systems; some have been implemented and evaluated for the first time in this
paper for possible use in Monte Carlo particle transport.
The analysis mainly concerns the energy range  
between 5 keV and a few MeV.
The validation process identifies the newly implemented model based on second
order S-matrix calculations as the one best reproducing experimental
measurements.
The validation results show that, along with Rayleigh scattering, additional
processes, not yet implemented in Geant4 nor in other major Monte Carlo systems,
should be taken into account to realistically describe photon elastic scattering
with matter above 1 MeV.
Evaluations of the computational performance of the various simulation algorithms
are reported along with the analysis of their physics capabilities.

\end{abstract}


\section{Introduction}
\label{sec_intro}
\PARstart{P}{hoton} elastic scattering is important in various experimental
domains, such as material analysis applications, medical diagnostics and imaging
\cite{cesareo1992}; more in general, elastic interactions contribute to the
determination of photon mass attenuation coefficients, which are widely used
parameters in medical physics and radiation protection \cite{hubbell2006}.
In the energy range between approximately 1~keV and few MeV, which is the object
of this paper, the resolution of modern detectors, high intensity synchrotron
radiation sources and, in recent years, the availability of resources for
large scale numerical computations have concurred to build a wide body of
knowledge on photon elastic scattering.
Extensive reviews of photon elastic scattering, that cover both its theoretical
and experimental aspects, can be found in the literature (e.g.
\cite{Kane1986,roy_kissel_pratt1999, bradley1999}).

This paper addresses this topic under a pragmatic perspective: its simulation in
general purpose Monte Carlo codes for particle transport.
Photon interactions with matter, both elastic and inelastic, play a critical
role in these systems; their modeling presents some peculiarities, because the
software must satisfy concurrent requirements of physical accuracy and
computational performance.

Photon-atom elastic scattering encompasses various interactions, but Rayleigh
scattering, i.e. scattering from bound electrons, is the dominant contribution
in the low energy r\'egime and, as energy increases, remains dominant in a
progressively smaller angular range of forward scattering.
Rayleigh scattering models are implemented in all general-purpose Monte
Carlo systems; comparison studies have highlighted discrepancies among some of
them \cite{emnist2005}, nevertheless a comprehensive, quantitative appraisal of
their validity is not yet documented in the literature.
It is worthwhile to note that the validation of simulation models implies their
comparison with experimental measurements \cite{trucano_what}; comparisons with
tabulations of theoretical calculations or analytical parameterizations, such as
those that are reported in \cite{cirrone2010} as validation of Geant4
\cite{g4nim,g4tns} photon cross sections, do not constitute a validation of the
simulation software.

%


This paper evaluates the models adopted by general-purpose Monte
Carlo systems and other modeling approaches not yet implemented
in these codes, to identify the state-of-the-art in the simulation of photon
elastic scattering.
Computational performance imposes constraints on the complexity of physics
calculations to be performed in the course of simulation: hence the analysis is
limited to theoretical models for which tabulations of pre-calculated values are
available, or that are expressed by means of simple analytical formulations.
To be relevant for general purpose Monte Carlo systems, tabulated data should
cover the whole periodic table of elements and an extended energy range.
The accuracy of elastic scattering simulation models is quantified with
statistical methods comparing them with a wide collection of experimental data
retrieved from the literature; the evaluation of physics capabilities is
complemented by the estimate of their computational performance.
These results provide guidance for the selection of physics models in simulation
applications in response to the requirements of physics accuracy and
computational speed pertinent to different experimental scenarios.

Special emphasis is devoted to the validation and possible improvement of photon
elastic scattering modeling in Geant4; nevertheless, the results documented in
this paper can be of more general interest also for other Monte Carlo systems.

\section{Physics overview}

Photon-atom elastic scattering has been the object of theoretical and
experimental interest for several decades; only a brief overview is summarized
here to facilitate the comprehension of the software developments and simulation
validation results documented in this paper.

Conventionally, photon-atom interactions are classified into elastic and
inelastic processes, depending on whether the photon energy changes in the
center of mass frame: although commonly applied in atomic physics practice and
Monte Carlo transport, this distinction is somewhat arbitrary, since QED
(quantum electrodynamics) radiative corrections and target recoil actually make
all processes inelastic and, on the experimental side, the classification of a
process as elastic or inelastic is limited by source bandwidth and detector
resolution.

The physics models considered in this study concern isolated
photon-atom scattering: they neglect the effects of the environment of the
target atom and multi-photon effects associated with incident beams.
They do not take into account either any internal degrees of freedom of the initial or
final atom, such as its orientation; therefore they do not describe 
X-ray scattering from electrons in molecules and oriented solids.
This scenario is consistent with the assumptions of general-purpose Monte Carlo
systems for particle transport, where particles are assumed to interact with
free atoms of the medium.


\subsection{Theoretical calculations}

%

Elastic photon-atom scattering encompasses various types of interactions, which
are usually distinguished as:
\begin{itemize}
\item Rayleigh scattering, for scattering from the atomic electrons,
\item nuclear Thomson scattering, for scattering from the nucleus
considered as a point charge,
\item Delbr\"uck scattering, for scattering from the field of the nucleus,
\item nuclear resonance scattering, for scattering from the internal structure
of the nucleus.
\end{itemize}
These amplitudes are coherent, i.e. not physically distinguishable; the total
elastic amplitude is commonly calculated as the sum of these separate
amplitudes.
There are still unresolved questions about the additivity of the amplitudes from
composite systems, due to the complex interplay in a composite system between
partitioning the total amplitude and the reduction from a many-body photon-atom
interaction to an effective single particle formalism \cite{pratt1994}; open issues
concern the effects of electron correlations and of the finite lifetimes of
atomic excited states \cite{Kane1986}.

Rayleigh scattering calculations are commonly performed by means of an
approximation, that describes the scattering amplitude in terms of the form
factor, which represents the Fourier transform of the charge density of the
atom.
This can be simply understood considering the two Thomson amplitudes from a free
(point-like) particle \cite{dyson}:
\begin{equation}
A^T_{||}=-r_0\cos\theta\,,\qquad A^T_{\perp}=-r_0
\end{equation}
leading to a differential cross section
\begin{equation}
\dfrac{d \sigma}{d \Omega} =\dfrac{r_{o}}{2} (1 + cos^{2} \theta) 
\label{eq_dcs}
\end{equation}
which, for coherent scattering, is modified in presence of a charge distribution in
\begin{equation}
\dfrac{d \sigma}{d \Omega} =\dfrac{r_{o}}{2} (1 + cos^{2} \theta) F^{2}(q,Z)
\label{eq_dcs2}
\end{equation}
where $\sigma$ represents the cross section, $\Omega$ is the solid angle,
$r_{0}$ is the classical electron radius, $\theta$ is the scattering angle,
$F(q,Z)$ is the atomic form factor as a function of momentum transfer $q$ and
atomic number Z of the target atom.
This approximation is valid for photon energies much greater than electron
binding energies and for non-relativistic momentum transfers, which in this case
is meant as $\hbar q\ll m_e c$, where $\hbar$ is the reduced Planck constant,
$m_e$ is the electron mass and $c$ is the speed of light.
In the so-called ``modified form factor'' approximation each subshell charge
density is multiplied by a different factor.

Various calculations of non-relativistic, relativistic and modified relativistic
form factors are documented in the literature; the tabulations by Hubbell et al.
\cite{hubbell1975}, Hubbell and {\O}verb{\o} \cite{overbo}, and Schaupp et al.
\cite{schaupp}, respectively of non-relativistic, relativistic and modified form
factors, are representative of this approach.

For forward scattering the difference between the form factor amplitude and the
exact amplitude is described by two anomalous scattering factors
\cite{Kane1986}, that take into account the persistence of binding effects even
in the high energy limit of Rayleigh scattering.

A more consistent approach for the description of elastic photon-atom scattering
involves the evaluation of the relativistic second order S-matrix element in
independent particle approximation.
Relativistic quantum electrodynamics, treated in lowest non-vanishing order in
$e^2$, provides the basic theoretical framework for these calculations; neglect
of higher order terms in $e^2$ in the calculation means that radiative
corrections are not taken into account.
The mass of the atomic nucleus is assumed to be infinite compared to the photon
energy in question; therefore only scattering from atomic electrons is
considered.
This many-electron state is treated in a self-consistent central field
approximation, in which each electron is assumed to move in an average potential
due to all atomic electrons and the nucleus.

Numerical evaluation of the second order S-matrix for single electron
transitions in a potential was first attempted in the 1950s by Brown and
co-workers \cite{brown1955}.
This calculation scheme requires considerable computing resources, since the
multipole expansion of the photon field converges slower and slower for
increasing energies; systematic evaluation of the S-matrix element has become
practical relatively recently, thanks to wide availability of large scale
computing facilities.
Extensive calculations based on the S-matrix approach (e.g.
\cite{kissel_pratt1985,pratt1994,kissel1995}) have been performed by Kissel,
Pratt and co-workers.


\subsection{Data libraries}

Some results of theoretical calculations of photon elastic scattering are publicly 
available in the form of tabulations distributed as data libraries.

EPDL97 \cite{epdl97} (EPDL Evaluated Photon Data Library, 1997 version), which
is part of the ENDF/B-VII.1 \cite{ENDFBVII} evaluated nuclear data file,
includes total cross sections, form factors and anomalous scattering factors for
atoms with atomic number between 1 and 100, and for photon energies from 1~eV to
100~GeV.
The form factors in EPDL97 are the non-relativistic ones calculated by Hubbell
\cite{hubbell1975}; the anomalous scattering factors are by Cullen
\cite{scatman}; the total cross sections derive from calculations combining
Thomson scattering, form factors and anomalous scattering factors, which were
numerically integrated.

EPDL97 documentation reports rough estimates and qualitative comments about the
accuracy of the tabulated data, but it does not document how these estimates
were produced.
To the best of our knowledge systematic, quantitative validation of EPDL97
coherent scattering data is not documented in the literature.

EPDL97 is extensively used in Monte Carlo simulation; details are given in section 
\ref{sec_mc}.

The RTAB \cite{rtab} database encompasses a set of tabulations of photon elastic
scattering cross sections, which are the result of various methods of
calculation, and of components for their calculations, such as form factors and
anomalous scattering factors.
The differential cross sections listed in RTAB are based on:
\begin{itemize}
\item numerical S-matrix calculations by Kissel and Pratt,
\item relativistic form factors,
\item non-relativistic form factors by Hubbell et al. \cite{hubbell1975},
\item modified relativistic form factors,
\item modified relativistic form factors with angle-independent anomalous scattering factors.
\end{itemize}

Apart from the data derived from other sources, such as Hubbell's form factors,
all the data in the RTAB database have been consistently computed
in the same Dirac-Slater potential.
The tabulations have been generated on a 97-point grid for scattering angles
between 0 and 180 degrees, and on a 56-point grid for energies between 54.3 eV
and 2.7541 MeV; they are listed for atomic numbers from 1 to 100.

Two sets of cross sections based on S-matrix tabulations are available in RTAB:
one takes into account both the Rayleigh and Thomson scattering amplitudes,
which are added coherently, and one is limited to the Rayleigh scattering
amplitude.
In the following, "S-matrix" or "SM" identify RTAB tabulations involving both
amplitudes; whenever tabulations limited to Rayleigh scattering are used, they
are explicitly identified in the text.

Comparisons of measured data with S-matrix calculations are reported in some
experimental publications, which are limited to the configuration (photon energy,
scattering angle and target material) of the experiment they describe.
A systematic, quantitative validation of RTAB data is not documented yet in the
literature.
This database has not been exploited yet in general purpose Monte Carlo systems.



\section{Photon elastic scattering in Monte Carlo codes}
\label{sec_mc}

General purpose Monte Carlo codes for particle transport include algorithms for
the simulation of Rayleigh scattering, most of which are based on the form factor 
approximation; they do not appear to account for other amplitudes involved in
photon elastic scattering, apart from the implicit inclusion of Thomson
scattering in the calculation of total cross sections by the codes that use
EPDL97.
 
These codes assume that photons interact with free atoms of the medium:
this assumption neglects that the molecular structure and the structure of the
medium can affect coherent scattering. 
Some codes provide users the option to directly input tabulated Rayleigh
scattering cross sections and form factors, which may account for molecular
effects specific to a given material;
such tabulations are available for a limited number of materials (e.g. \cite{morin1982}).
Photon elastic scattering by molecules is not treated in this paper.
The information summarized here concerns models implemented in Monte Carlo codes
to describe elastic scattering by non-polarized photons; some Monte Carlo
systems also include algorithms for Rayleigh scattering by polarized photons,
which are not considered here.



EGS5 \cite{egs5} and EGS4 \cite{egs4} calculate total Rayleigh scattering cross
sections from the tabulations by Storm and Israel \cite{storm_israel_photon},
which in turn derive from the integration over angle of equation (\ref{eq_dcs}),
where form factors by Hanson \cite{hanson} were used.
They sample the coherent scattering angle based on the relativistic form factors
by Hubbell and {\O}verb{\o} \cite{overbo}.
EGSnrc \cite{egsnrc} can use three sets of cross section data: those by Storm
and Israel as EGS4, and the additional options of EPDL97 and XCOM \cite{xcom}
cross sections.

ETRAN \cite{etran} uses the cross sections of the XCOM database, which are based
on the relativistic form factors of Hubbell and {\O}verb{\o} \cite{overbo}, and
samples the change in photon direction according to the form factor
approximation.

FLUKA \cite{fluka1,fluka2} simulates Rayleigh scattering based on EPDL97.

ITS \cite{its5} calculates total Rayleigh scattering cross sections based on the
relativistic form factors by Hubbell and {\O}verb{\o} \cite{overbo}, but it
samples the scattering angle based on the non-relativistic form factors by
Hubbell et al. \cite{hubbell1975}.

MCNP5 \cite{mcnp5} and MCNPX \cite{mcnpx} neglect Rayleigh scattering 
when the ``simple option'' of photon transport is chosen; otherwise they
also base the simulation on the form factor approximation. 
They provide different options of Rayleigh scattering data, the most recent of
which uses EPDL97; the other options use the older EPDL89 \cite{epdl89} version
of EPDL, data from ENDF/B-IV \cite{ENDFB-IV} based on non-relativistic form
factors by Hubbell et al. \cite{hubbell1975}, and the tabulations by Storm and
Israel \cite{storm_israel_photon}, the latter limited to a few elements with
atomic number greater than 83.
Nuclear resonance fluorescence, nuclear Thomson scattering and Delbr\"uck
scattering are not treated \cite{mcnpx27e}.

Penelope \cite{penelope} 2008 \cite{penelope2008} and 2011 \cite{penelope2011}
versions calculate total Rayleigh scattering cross sections and scattering
angles based on EPDL97 \cite{epdl97}; earlier versions exploited analytical
approximations \cite{baro1994a} to Hubbell's non-relativistic form factors for
the calculation of differential and total cross sections.
While Penelope 2008 and 2011 documentation \cite{penelope2008,penelope2011}
states that the total cross sections and form factors used in Penelope are from
EPDL97, the tabulations of these quantities distributed with the Penelope code
appear different from those in EPDL97: presumably, the tabulations included in
Penelope have been recalculated, or interpolated, over a different energy or
momentum transfer grid.

GEANT 3 \cite{geant3} handled Rayleigh scattering according to empirical
formulae derived from EGS3 \cite{egs3}; they consist of polynomial fits to cross
sections by Storm and Israel, and relativistic form factors by Hubbell and
{\O}verb{\o} \cite{overbo}.

The Geant4 toolkit encompasses various implementations of Rayleigh scattering.
The latest versions at the time of writing this paper are Geant4 9.4p03 and
Geant4 9.5, both released in December 2011 (9.4p03 being the most recent).
The Geant4 \textit{G4OpRayleigh} class implements a Rayleigh scattering model
only applicable to a particle type identified in Geant4 as ``optical photon'',
which is an object of the \textit{G4OpticalPhoton} class; this process is not
considered in the following analysis, that deals with Geant4 Rayleigh
scattering models concerning ordinary photons, which are instances of the
\textit{G4Gamma} class.

Functionality for the simulation of Rayleigh scattering of ordinary photons was
first introduced \cite{lowe_e} in the low energy electromagnetic package
\cite{lowe_chep,lowe_nss} of Geant4 0.1 version, based on total cross sections
and non-relativistic form factors tabulated in the EPDL97 \cite{epdl97} data
library; this modeling approach is identified in Geant4 9.5 as ``Livermore
Rayleigh model''.

Geant4 also includes two implementations of Rayleigh scattering reengineered
from the Penelope \cite{penelope} Monte Carlo code, respectively
equivalent to the Rayleigh scattering algorithms in Penelope 2001
\cite{penelope2001} and 2008 \cite{penelope2008} versions, the latter
also based on EPDL97  as the original Rayleigh scattering implementation
available in Geant4.
The reengineered Penelope 2008 code uses the tabulations of Rayleigh
scattering total cross sections and form factors distributed with Penelope,
which, as discussed above, appear different from the native EPDL97 values used
by the ``Livermore Rayleigh'' implementation.


A further Rayleigh scattering simulation model, implemented in the
\textit{G4XrayRayleighModel} class, has been first released in Geant4 9.5
version: it is defined in the software release notes as ``implementing
simplified Rayleigh scattering'', but no further information could be found in
Geant4 documentation, nor in the literature, about the physics model underlying
this implementation.
From the C++ source code, the authors of this paper could evince that this
model calculates the total Rayleigh scattering cross section according to an
analytical formula involving numerical parameters, and samples the scattering
angle according to the distribution of Thomson scattering from a point-like
charge.


\section{Strategy of this study}

An extensive set of simulation models, which are representative of the variety
of theoretical approaches documented in the literature, have been evaluated to
identify the state-of-the-art of photon elastic scattering in the context of
Monte Carlo particle transport.
The physics models considered in this analysis involve the implementation of simple
formulae in the simulation software or exploit available tabulations of complex
theoretical calculations.


The models for photon elastic scattering simulation evaluated in this paper concern
total and differential cross sections: in particle transport, the former are
relevant to determine the occurrence of the scattering process, while the latter
determine the actual outcome of the scattering by defining the angular
distribution of the scattered photon at a given energy and for a given target.

All the models subject to study have been implemented in a consistent software
design, compatible with the Geant4 toolkit, which minimizes external
dependencies to ensure the unbiased appraisal of their intrinsic capabilities.

A wide set of experimental data of photon elastic scattering has been collected
from the literature for this study; simulation models are validated through comparison
with these measurements.
The compatibility with experiment for each model, and the differences in
compatibility with experiment across the various models, are quantified by means
of statistical methods.

The measurements of photon elastic scattering reported in the literature mainly
concern differential cross sections; therefore the validation process is focused
on this observable.
Due to the scarcity of total cross section measurements in the literature, the
capability of the simulation models to calculate total cross sections consistent
with experiment can be directly tested only over a small data sample; 
nevertheless, it can be indirectly inferred for total cross sections deriving
from the integration of validated differential cross sections.

The validation of the physics capabilities of the simulation models is
complemented by the evaluation of their computational performance.


\subsection{Simulation models}


\begin{table*}[htbp]
  \centering
  \caption{Simulation models of photon elastic scattering: differential cross sections}
    \begin{tabular}{|l|l|c|}
    \hline
    \textbf{Identifier} & \textbf{Underlying theoretical approach} & \textbf{Tabulated data source} \\
\hline
EPDL & Non-relativistic form factors & EPDL97 \\ 
EPDLASF & Non-relativistic form factors + anomalous scattering factors & EPDL97 \\
Penelope 2001 & Parameterization of  non-relativistic form factors (as in Penelope 2001) & - \\
Penelope 2008  &  Non-relativistic form factors (as in Penelope 2008-2011) & Retabulated EPDL97 \\
RF & Relativistic form factors & RTAB \\
NF & Non-relativistic form factors  & Hubbell 1975, as reported in RTAB\\
MF & Modified form factors  & RTAB \\
RFASF & Relativistic form factors and anomalous scattering factors  & RTAB \\
MFASF & Modified form factors and anomalous scattering factors  & RTAB \\
SM & Second order S-matrix & RTAB \\
    \hline
    \end{tabular}%
  \label{tab_models}%
\end{table*}%

The simulation models of differential cross section analyzed in this study 
implement calculations based on second order S-matrix and on the form
factor approximation; they are summarized in Table \ref{tab_models}.
Some of these simulation models represent novel approaches with respect to those
so far available in Geant4 and in other general purpose Monte Carlo codes;
among them, the paper examines whether a model based on S-matrix calculations,
which constitutes a more consistent theoretical approach, but requires more
complex computational operations, would be sustainable in the context of Monte
Carlo particle transport.



 
Various simulation models are based on the non-relativistic form factors by
Hubbell et al. \cite{hubbell1975}: they are identified in Table~\ref{tab_models}
as ``EPDL'', ``NF'', ``Penelope 2001'' and ``Penelope 2008''.
They use different tabulations of the form factors as a function of
momentum transfer, or different algorithms to sample the scattered photon
direction based on the form factors.
The model identified as ``EPDL'' uses non relativistic form factors as they are
tabulated in EPDL97; it corresponds to the original Rayleigh scattering
simulation method in Geant4 low energy electromagnetic package \cite{lowe_e},
which has been reengineered in the context of the policy-based class design
described in section \ref{sec_sw}.
The models identified in this paper as ``Penelope 2001'' and ``Penelope 2008''
correspond to algorithms reengineered from the respective Penelope
versions: as mentioned in section \ref{sec_mc}, the Penelope 2008 model is based
on the EPDL97 data library, although re-tabulated over a different grid, while
the Penelope 2001 model exploits analytical approximations to Hubbell's
non-relativistic form factors.
The model identified in this paper as ``NF'' exploits the non-relativistic form
factors listed in RTAB.

Models based on relativistic and modified form factors, based on RTAB
tabulations, are respectively identified as ``RF'' and ``MF''.
To the best of the authors' knowledge, modified form factors have not been
used yet in general purpose Monte Carlo codes.

Additionally, calculations accounting for anomalous scattering factors along with
form factors are evaluated: the ``RFASF'' and ``MFASF'' models are based on RTAB
tabulations, while the ``EPDLASF'' model exploits the anomalous scattering
factors included in EPDL97, which have not been used yet in any of the
general-purpose Monte Carlo systems relying on EPDL97.

The model identified as ``SM'' is based on RTAB tabulations of second order 
S-matrix calculations.
This model accounts for the sum of two coherent amplitudes: the so-called 
Rayleigh amplitude and nuclear Thomson scattering.
A model based on S-matrix calculations limited to the Rayleigh scattering
amplitude is discussed in section \ref{sec_nuclear}.
To the best of the authors' knowledge, this paper documents the first
implementation based on S-matrix calculations for the simulation of photon elastic 
scattering by a general-purpose Monte Carlo system.

Total cross sections are calculated corresponding to all the models listed in
Table \ref{tab_models}; those that are highlighted as most relevant
by the validation analysis of differential cross sections in section
\ref{sec_diffcs} are discussed in detail in section \ref{sec_totcs}.
Furthermore, the total cross sections calculated by interpolation of Storm and
Israel \cite{storm_israel_photon} tables, by the XCOM Photon Cross Section
Database distributed by NIST (National Institute of Standards), by Brennan and
Cowan \cite{Brennan1992} parameterizations of McMaster's \cite{McMaster1969}
cross sections (commonly used in photon science), and by Geant4
\textit{G4XrayRayleighModel} are evaluated. 
The energy ranges covered by these compilations are respectively 1-100~MeV (Storm and Israel), 
0.03-695~keV (Brennan and Cowan) and 1 keV to 100 GeV (XCOM).
The total cross section models examined in this paper are summarized in
Table~\ref{tab_totcs}.

\begin{table*}[htbp]
  \centering
  \caption{Simulation models of photon elastic scattering: total cross sections}
    \begin{tabular}{|l|l|c|}
    \hline
    \textbf{Identifier} & \textbf{Method} & \textbf{Source} \\
\hline
EPDL & Non-relativistic form factors with anomalous scattering factors & EPDL97 \\ 
Penelope 2001 & Analytical model & Penelope 2001 \\
MFASF & Integration of differential cross sections based on modified form factors with anomalous scattering factors  & RTAB \\
SM & Integration of cross sections based on second order S-matrix & RTAB \\
XCOM & Based on form factor approximation & NIST \\
Storm & Storm and Israel tabulations & \cite{storm_israel_photon} \\
Brennan &  Brennan and Cowan's parameterization of McMaster et al. cross sections & \cite{Brennan1992} \\
G4std & Parameterization in the Geant4 \textit{G4XrayRayleighModel} class & Not documented \\
    \hline
    \end{tabular}%
  \label{tab_totcs}%
\end{table*}%

Tabulations of total Rayleigh scattering cross sections are directly available
in EPDL97, XCOM and \cite{storm_israel_photon}; tabulations for other
physics models have been produced on the same energy grid as EPDL97 to 
facilitate the comparison of the various models.
The tabulations associated with the simulation models based on RTAB listed in Table
\ref{tab_models} were produced by integrating the corresponding differential
cross sections over the angle.

%

\subsection{Software environment}
\label{sec_sw}

All the physics models evaluated in this paper have been implemented in the same
software environment, which is compatible with Geant4; computational features
specific to the original physics algorithms have been preserved as much as possible. 
The uniform software configuration ensures an unbiased appraisal of the
intrinsic characteristics of the various physics models.
The correctness of implementation has been verified prior to the validation 
process to ensure that the software reproduces the physical features
of each model consistently.

The software adopts a policy-based class design \cite{alexandrescu}; this 
technique was first introduced in a general-purpose Monte Carlo system in
\cite{tns_dna}.
This programming paradigm supports the provision of a series of alternative
physics models for Monte Carlo transport,  characterized by high granularity,
which can be used interchangeably with great versatility, without imposing the
burden of inheritance from a pre-defined interface, since policies are
syntax-oriented, rather than signature-oriented.


Two policies have been defined for the simulation of photon elastic scattering,
corresponding to the calculation of total cross section and to the generation of
the scattered photon; they conform to the prototype design described in
\cite{em_chep2009,em_nss2009}.
A photon elastic scattering process, derived from the \textit{G4VDiscreteProcess}
class of Geant4 kernel, acts as a host class for these policy classes.
All the simulation models implemented according to this policy-based class design
are compatible for use with Geant4, since Geant4 tracking handles 
all discrete processes polymorphically through the \textit{G4VDiscreteProcess} 
base class interface.

A single policy class calculates total cross sections for all the physics models
that exploit tabulations; alternative tabulations, corresponding to different
physics models, are managed through the file system.
Specific policy classes implement the analytical calculations of total cross sections
as in Penelope 2001 and \textit{G4XrayRayleighModel}. 

Two alternative policy classes sample the direction of the scattered photon
utilizing the form factor approximation: they differ by the sampling algorithm,
respectively based on the inverse transform method and on acceptance-rejection
sampling \cite{mc_rubinstein}.
Both can be used with any tabulated form factors, which are managed through the
file system.

A dedicated policy class deals with the generation of the scattered
photon based on S-matrix calculations; the scattering angle is sampled
based on the inverse transform sampling method.

Dedicated policy classes generate the scattered photon according to the
analytical formulations used in Penelope 2001 and \textit{G4XrayRayleighModel}.

The final state generation based on the form factor approximation involves
one-dimensional interpolations only, while the use of S-matrix calculations
requires bi-dimensional interpolation of RTAB data.
Interpolation algorithms are discussed in section \ref{sec_interpol}.

The software design adopted in this study ensures greater flexibility than the
design currently adopted in Geant4 electromagnetic package, since it allows
independent modeling (and test) of total cross section calculation and photon
scattering generation.
Since policy classes are characterized by a single responsibility and have
minimal dependencies on other parts of the software, the programming paradigm
adopted in this study also facilitates the validation of physics models:
validation tests involve only the instantiation of an object of the policy class and
the invocation of the member function implementing the policy, therefore they
expose only the intrinsic features of the physics models, excluding possible
effects on physics behaviour due to other parts of the Monte Carlo system.
Equivalent validation tests of physics models as currently implemented in Geant4
require instead a full scale simulation application, including the creation of a
geometry model even when it is conceptually redundant (e.g. to validate a cross
section calculation).

\subsection{Experimental data}
\label{sec_exp}

Experimental data for the validation of the simulation models were collected
from a survey of the literature.

The sample of experimental differential cross sections consists of approximately
4500 measurements \cite{Kane1986},\cite{Alvarez1958}-\cite{Wilson1953}, which
concern 69 target atoms and span energies from 5.41~keV to 39~MeV, and
scattering angles from 0.5 to 165 degrees.
An overview of this data sample is summarized in Tables \ref{tab_exp1} and
\ref{tab_exp2}.

Experimental total cross section data are scarce in the literature.
To the best of our efforts we could retrieve only a journal publication
\cite{Gowda1995} reporting a measurement at 661.6 keV for four elements: barium,
tungsten, lead and bismuth.
A larger set of measurements by the same group as \cite{Gowda1995}, concerning
18 target elements and two photon energies (279.2~keV and 661.6~keV), is
documented in \cite{Gowda_thesis}; this reference includes the four values
published in \cite{Gowda1995}.
The experimental errors reported in \cite{Gowda_thesis} vary between 3\% and 6\%
for the measurements at 661.6 keV, and between 0.3\% and 0.5\% for the data
at 279.2 keV: since the reported precision of the two sets of measurements in
similar experimental conditions differs by an order of magnitude, and better
than 1\% precision appears inconsistent with typical experimental errors in the
field documented in the literature, one may wonder whether some of the
experimental uncertainties listed in \cite{Gowda_thesis} could be affected by a
typographical slip.

It is worthwhile to note that the available experimental differential cross section data 
are too scarce to calculate total cross sections based on them.
Therefore they could not be exploited for the validation of theoretical total cross section models.

A small sample of experimental measurements of photon elastic scattering at the
electronvolt scale has been retrieved; data at such low energy are scarce in the
literature, and consist mostly of measurements of molecular cross sections,
which are outside the scope of this paper.
A few measurements of absolute total cross sections, that concern atomic
gaseous targets (mostly noble gases) in the energy range of
approximately 1-10~eV, have been retrieved in the literature:
the intrinsic physical characteristics of these data are congruous for
comparison with calculations performed in independent particle approximation,
since molecular effects, which become important at low energies, are expected to
be minimized in atomic gaseous targets.
Some of these data involved normalizations to other values to obtain absolute
cross sections: these manipulations can be source of systematic effects and,
if the normalization is based on theoretical references, would make the
data inappropriate for the validation of simulation models.
Although this very low energy data sample is too limited to serve for proper
simulation validation, it enables at least a qualitative appraisal of EPDL97
cross sections, that extend down to 1~eV.



\begin{table*}
\begin{center}
\caption{Summary of the experimental data used in the validation analysis: atomic numbers 3-72}
\label{tab_exp1}
\begin{tabular}{llcccp{10cm}}\hline
\multicolumn{2}{c}{Element} & Energy range & Angle range & Sample  & References \\ 
$Z$  &  Symbol               & (keV)    & (degrees)  & size         &            \\ 
\hline 
3 & Li & 1600 & 124 & 1 & \cite{Alvarez1958} \\ 
4 & Be & 59.54 & 45 & 9 & \cite{Icelli2001} \\ 
5 & B & 145.4 & 2.03 & 4 & \cite{Gupta1979,Gaspar1986} \\ 
6 & C & 22.1--145.4 & 2.03--133 & 13 & \cite{Gupta1979,Roy1975,Gaspar1986,Grag1993,Kumar2002} \\ 
7 & N & 145.4 & 2.03 & 1 & \cite{Gupta1979} \\ 
12 & Mg & 22.16--59.54 & 90--121 & 4 & \cite{Rao1996,Shahi1998,Shahi1997} \\ 
13 & Al & 14.93--1600 & 1.6--145 & 86 & \cite{Rao1996,Casnati1990,Eichler1985,Alvarez1958,Basavaraju1995a,Basavaraju1995b,Varier1989,Kane1987,Puri1996,Shahi1998,Shahi1997,Standing1962,Roy1975,Schopper1957,Gaspar1986,Guy1992,Gowda1986,Icelli2001,Kane1994,Rao1994,Rao1996a,Rao1996b,Rao1997a,Rao1995} \\ 
14 & Si & 17.44--59.54 & 90--121 & 3 & \cite{Shahi1998,Grag1993,Rao2000a} \\ 
16 & S & 22.1--59.54 & 121--133 & 2 & \cite{Shahi1998,Kumar2002} \\ 
21 & Sc & 17.44--30.85 & 90--90 & 2 & \cite{Rao1997a,Rao2000a} \\ 
22 & Ti & 17.44--59.54 & 10--160 & 18 & \cite{Bui1989,Mandal2002,Puri1996,Shahi1998,Shahi1997,Kumar2009,Rao1997a,Rao2000a} \\ 
23 & V & 17.44--59.54 & 10--160 & 17 & \cite{Casnati1990,Mandal2002,Shahi1998,Kumar2009,Grag1993,Rao1997a,Rao2000a} \\ 
24 & Cr & 17.44--33.29 & 90--90 & 9 & \cite{Rao1996,Rao1996a,Rao1997a,Rao2000a} \\ 
25 & Mn & 17.44--33.29 & 90--90 & 4 & \cite{Rao1996,Rao1997a,Rao2000a} \\ 
26 & Fe & 6.4--1120.5 & 2.72--130 & 38 & \cite{Cindro1958,Bui1989,Mandal2002,Puri1996,Shahi1998,Shahi1997,Roy1975,Erzeneoglu1997,Grag1993,Rao1996a,Rao1997a,Rao2000a} \\ 
27 & Co & 17.44--59.54 & 10--160 & 13 & \cite{Puri1996,Shahi1997,Kumar2009,Rao1997a,Rao2000a} \\ 
28 & Ni & 7.47--81 & 45--145 & 22 & \cite{Rao1996,Basavaraju1995a,Bui1989,Mandal2002,Puri1996,Shahi1998,Shahi1997,Kane1994,Rao1997a,Rao2000a} \\ 
29 & Cu & 8.04--1408.1 & 1.02--165 & 306 & \cite{Cindro1958,Rao1996,Chitwattanagorn1987,Bradley1990,Eichler1985,Kane1983,Barros1981,Bui1989,Eichler1983,Elyaseery1998,Goncalves2000,Kahane1996,Varier1989,Lestone1988,Mandal2002,Puri1996,Shahi1998,Shahi1997,Gupta1979,Ramanathan1979,Standing1962,Barros1981b,Nath1964,Roy1975,Schopper1957,Nayak1992,Siddappa1989,Gaspar1986,Guy1992,Nayak1993,Gowda1986,Elyaseery1998a,Grag1993,Rao1994,Rao1996a,Rao1996b,Rao1997a,Rao2000a,Rao1995,Elyaseery2000} \\ 
30 & Zn & 8.63--661.6 & 10--165 & 87 & \cite{Schumacher1977,Smend1973,Eichler1985,Bui1989,Elyaseery1998,Goncalves1986,Mandal2002,Puri1996,Shahi1998,Shahi1997,Kumar2009,Kumar2001,Elyaseery1998a,Erzeneoglu1997,Grag1993,Rao1997a,Rao2000a,Chatterjee1998,Elyaseery2000} \\ 
31 & Ga & 16.58 & 90 & 1 & \cite{Rao1996} \\ 
32 & Ge & 9.8--59.54 & 90--133 & 4 & \cite{Bui1989,Shahi1998,Kumar2002} \\ 
33 & As & 22.1 & 133 & 1 & \cite{Kumar2002} \\ 
34 & Se & 22.1 & 133 & 1 & \cite{Kumar2002} \\ 
36 & Kr & 59.54 & 20 & 11 & \cite{Smend1986} \\ 
37 & Rb & 22.1 & 133 & 1 & \cite{Kumar2002} \\ 
38 & Sr & 14.14--30.85 & 90--90 & 10 & \cite{Bui1989,Rao1994,Rao1996a,Rao1997a} \\ 
39 & Y & 17.44--59.54 & 10--160 & 19 & \cite{Rao1996,Shahi1998,Shahi1997,Kumar2009,Rao1996b,Rao1997a,Rao2000a,Rao1995} \\ 
40 & Zr & 13.95--661.6 & 10--165 & 67 & \cite{Bui1989,Elyaseery1998,Mandal2002,Puri1996,Shahi1998,Shahi1997,Murty1965,Kumar2009,Prasad1978,Chong1996,Elyaseery1998a,Prasad1978a,Rao1996a,Rao1997a,Elyaseery2000} \\ 
41 & Nb & 13.95--59.54 & 10--165 & 66 & \cite{Bui1989,Elyaseery1998,Mandal2002,Puri1996,Shahi1998,Shahi1997,Kumar2009,Chong1996,Elyaseery1998a,Erzeneoglu1997,Rao1997a,Rao2000a,Elyaseery2000} \\ 
42 & Mo & 13.95--1408.1 & 2--165 & 220 & \cite{Schumacher1977,Rao1996,Smend1973,Basavaraju1979,Casnati1990,Barros1981,Bui1989,Elyaseery1998,Varier1989,Lestone1988,Mandal2002,Puri1996,Shahi1998,Shahi1997,Barros1981b,Kumar2009,Nandi1989,Kumar2001,Nayak1992,Siddappa1989,Guy1992,Nayak1993,Chong1996,Elyaseery1998a,Grag1993,Rao1997a,Rao1995,Taylor1987,Elyaseery2000} \\ 
44 & Ru & 22.16--59.54 & 90--121 & 2 & \cite{Bui1989,Shahi1998} \\ 
45 & Rh & 22.16--59.54 & 117--121 & 2 & \cite{Shahi1998,Shahi1997} \\ 
46 & Pd & 5.41--59.54 & 90--130 & 18 & \cite{Rao1996,Mandal2002,Puri1996,Shahi1998,Shahi1997,Rao1996a,Rao1997a,Rao1998,Rao2000a} \\ 
47 & Ag & 5.41--661.6 & 5.12--165 & 125 & \cite{Rao1996,Barros1980,Barros1981,Anand1965,Bui1989,Elyaseery1998,Goncalves2000,Shahi1998,Barros1981b,Kumar2009,Anand1963,Nayak1992,Siddappa1989,Nayak1993,Elyaseery1998a,Rao1996a,Rao1997a,Rao1998,Rao2000a,Kumar2002,Elyaseery2000} \\ 
48 & Cd & 5.41--1600 & 2.4--165 & 173 & \cite{Rao1996,Barros1980,Casnati1990,Eichler1985,Barros1981,Alvarez1958,Bui1989,Eichler1983,Elyaseery1998,Goncalves2000,Mandal2002,Puri1996,Shahi1998,Shahi1997,Ramanathan1979,Barros1981b,Kumar2009,Nandi1989,Nayak1992,Siddappa1989,Gaspar1986,Nayak1993,Elyaseery1998a,Rao1994,Rao1997a,Rao1997b,Rao1998,Rao2000a,Elyaseery2000} \\ 
49 & In & 5.41--1302 & 1.02--165 & 63 & \cite{Rao1996,Bui1989,Elyaseery1998,Kahane1996,Mandal2002,Puri1996,Shahi1998,Shahi1997,Elyaseery1998a,Rao1996b,Rao1997a,Rao1997b,Rao1998,Rao2000a,Elyaseery2000} \\ 
50 & Sn & 5.41--1408.1 & 1.03--165 & 405 & \cite{Cindro1958,Schumacher1977,Rao1996,Murty1965b,Smend1973,Bradley1986,Bradley1990,Casnati1990,Kane1983,Barros1981,Bernstein1958,Banaigs1958,Anand1965,Bui1989,Elyaseery1998,Goldzahl1957,Goncalves1986,Hara1958,Lestone1988,Mandal2002,Puri1996,Shahi1998,Shahi1997,Gupta1979,Murty1965,Standing1962,Barros1981b,Kumar2009,Nandi1989,Prasad1978,Nath1964,Anand1963,Roy1975,Bradley1989,Kumar2001,Nayak1992,Siddappa1989,Mann1956,Guy1992,Nayak1993,Gowda1986,Elyaseery1998a,Prasad1978a,Rao1997a,Rao1997b,Rao1998,Rao2000a,Taylor1987,Elyaseery2000} \\ 
51 & Sb & 5.41--59.54 & 90--131 & 24 & \cite{Rao1996,Bui1989,Shahi1998,Rao1997b,Rao1998,Rao2000a} \\ 
52 & Te & 22.1--59.54 & 121--133 & 2 & \cite{Shahi1998,Kumar2002} \\ 
53 & I & 661.6--1120.5 & 95--110 & 4 & \cite{Cindro1958,Anand1963} \\ 
54 & Xe & 59.54--661.6 & 3.02--120 & 25 & \cite{Smend1986,Goncalves1999} \\ 
55 & Cs & 17.44 & 90 & 1 & \cite{Rao2000a} \\ 
56 & Ba & 17.44--1120.5 & 57.5--133 & 15 & \cite{Cindro1958,Barros1980,Bui1989,Shahi1998,Kumar2001,Rao1997a,Rao2000a,Kumar2002} \\ 
57 & La & 17.44--661.6 & 1.77--90 & 3 & \cite{Gupta1979,Rao1997a,Rao2000a} \\ 
58 & Ce & 14.93--59.54 & 90--131 & 6 & \cite{Bui1989,Rao1994,Rao1997a,Rao2000a} \\ 
59 & Pr & 14.93--279.2 & 90--90 & 10 & \cite{Nayak1992,Siddappa1989,Nayak1993,Rao1994,Rao1996,Rao1997a,Rao2000a} \\ 
60 & Nd & 22.16--661.6 & 1.77--135 & 11 & \cite{Smend1973,Bui1989,Gupta1979} \\ 
62 & Sm & 14.93--279.2 & 90--133 & 13 & \cite{Bui1989,Mandal2002,Nayak1992,Siddappa1989,Nayak1993,Rao1994,Rao1996,Rao1997a,Kumar2002} \\ 
63 & Eu & 17.44 & 90 & 1 & \cite{Rao2000a} \\ 
64 & Gd & 17.44--320 & 10--160 & 34 & \cite{Bui1989,Mandal2002,Shahi1998,Shahi1997,Rao1991,Kumar2009,Baraldi1996,Kumar2001,Nayak1992,Siddappa1989,Nayak1993,Rao1997a,Rao2000a} \\ 
65 & Er & 59.54--320 & 45--121 & 11 & \cite{Shahi1998,Rao1991} \\ 
66 & Dy & 16.58--661.6 & 1.77--160 & 36 & \cite{Bui1989,Mandal2002,Puri1996,Shahi1998,Shahi1997,Gupta1979,Rao1991,Kumar2009,Baraldi1996,Nayak1992,Siddappa1989,Nayak1993,Rao1996,Rao1997a,Rao2000a} \\ 
67 & Ho & 16.58--279.2 & 10--160 & 20 & \cite{Bui1989,Shahi1998,Kumar2009,Kumar2001,Nayak1992,Siddappa1989,Nayak1993,Rao1996,Rao1997a,Rao2000a,Kumar2002} \\ 
68 & Er & 16.58--59.54 & 10--165 & 22 & \cite{Mandal2002,Shahi1998,Shahi1997,Kumar2009,Nandi1989,Baraldi1996,Rao1996,Rao1997a,Rao2000a} \\ 
69 & Tm & 22.1 & 133 & 1 & \cite{Kumar2002} \\ 
70 & Yb & 16.58--320 & 10--160 & 35 & \cite{Bui1989,Varier1989,Mandal2002,Shahi1998,Rao1991,Kumar2009,Baraldi1996,Kumar2001,Nayak1992,Siddappa1989,Nayak1993,Rao1996,Rao1997a,Rao2000a,Kumar2002} \\ 
72 & Hf & 22.16--511 & 0.5--160 & 13 & \cite{Bui1989,Hauser1966,Kumar2009,Baraldi1996} \\ 
\hline 
\end{tabular}
\end{center}
\end{table*}

\begin{table*}
\begin{center}
\caption{Summary of the experimental data used in the validation analysis: atomic numbers 73-92}
\label{tab_exp2}
\begin{tabular}{llcccp{10cm}}\hline
\multicolumn{2}{c}{Element} & Energy range & Angle range & Sample  & References \\ 
$Z$  &  Symbol               & (keV)    & (degrees)  & size         &            \\ 
\hline 
73 & Ta & 13.95-1408.1 & 0.5-165 & 295 & \cite{Schumacher1977,Murty1965b,Smend1973,Basavaraju1979,Basavaraju1995a,Bui1989,Elyaseery1998,Kahane1996,Varier1989,Lestone1988,Puri1996,Hauser1966,Shahi1998,Shahi1997,Murty1965,Ramanathan1979,Kumar2009,Prasad1978,Baraldi1996,Kumar2001,Nayak1992,Siddappa1989,Guy1992,Nayak1993,Elyaseery1998a,Prasad1978a,Kane1994,Taylor1987,Chatterjee1998,Elyaseery2000} \\ 
74 & W & 13.95-1408.1 & 0.5-165 & 151 & \cite{Chitwattanagorn1987,Taylor1981,Barros1980,Barros1981,Anand1965,Elyaseery1998,Puri1996,Hauser1966,Shahi1998,Shahi1997,Barros1981b,Kumar2009,Anand1963,Kumar2001,Nayak1992,Siddappa1989,Nayak1993,Elyaseery1998a,Elyaseery2000} \\ 
75 & Re & 22.16-511 & 0.5-121 & 3 & \cite{Hauser1966,Shahi1998,Shahi1997} \\ 
77 & Ir & 22.16-511 & 0.5-160 & 11 & \cite{Hauser1966,Shahi1998,Shahi1997,Kumar2009} \\ 
78 & Pt & 5.41-1120.5 & 0.5-160 & 100 & \cite{Murty1965b,Barros1980,Goncalves2000,Puri1996,Hauser1966,Shahi1998,Shahi1997,
Murty1965,Kumar2009,Kumar2001,Rao1994,Rao1996,Rao1997a,Rao1998a,Rao1998,Rao2000a} \\ 
79 & Au & 5.41-511 & 0.5-160 & 108 & \cite{Schumacher1977,Rao1996,Basavaraju1995a,Basavaraju1995b,Bui1989,Varier1989,Kane1987,Mandal2002,Hauser1966,Shahi1998,Shahi1997,Gupta1982,Kumar2009,Kumar2001,Erzeneoglu1996,Kane1994,Grag1993,Rao1994,Rao1996a,Rao1996b,Rao1997a,Rao1998a,Rao1998,Rao2000a,Rao1995} \\ 
80 & Hg & 22.1-1600 & 0.5-135 & 16 & \cite{Basavaraju1970,Cindro1958,Schumacher1973,Alvarez1958,Hauser1966,Shahi1998,Gupta1979,Anand1963,Kumar2001,Kumar2002} \\ 
81 & Tl & 22.1-511 & 0.5-133 & 3 & \cite{Hauser1966,Kumar2001,Kumar2002} \\ 
82 & Pb & 5.41-1613 & 0.5-160 & 797 & \cite{Basavaraju1970,Cindro1958,Kane1986,Schumacher1973,Schumacher1977,Rao1996,Murty1965b,Smend1973,
Chitwattanagorn1987,Basavaraju1979,Bradley1986,Bradley1990,Casnati1990,Hardie1970,Hardie1971,Eichler1985,Schumacher1969,
Chitwattanagorn1980,Kane1983,Messelt1956,Barros1981,Bernstein1958,Banaigs1958,Anand1965,Basavaraju1995a,Basavaraju1995b,
Bui1989,Eichler1983,Goldzahl1957,Goncalves1986,Goncalves2000,Hara1958,Kahane1992,Kane1978,Varier1989,Kane1987,Kasten1986,
Lestone1988,Mandal2002,Puri1996,Hauser1966,Shahi1998,Shahi1997,Gupta1982,Gupta1979,Murty1965,Ramanathan1979,Dixon1968,
Standing1962,Barros1981b,Kumar2009,Prasad1978,Nath1964,Wilson1953,Anand1963,Quivy1967,Schopper1957,Bradley1989,Kumar2001,
Nayak1992,Siddappa1989,Mann1956,Gaspar1986,Guy1992,Nayak1993,Gowda1986,Erzeneoglu1996,Prasad1978a,Kane1994,Grag1993,
Rao1994,Rao1996a,Rao1996b,Rao1997a,Rao1998a,Rao1998,Rao2000a,Rao1995,Taylor1987,Chatterjee1998} \\ 
83 & Bi & 22.16-511 & 0.5-131 & 9 & \cite{Basavaraju1995b,Bui1989,Kane1987,Hauser1966,Shahi1998,Shahi1997,Kumar2001} \\ 
90 & Th & 22.16-661.6 & 0.5-125 & 11 & \cite{Hauser1966,Shahi1998,Shahi1997,Gupta1979,Kumar2001,Nayak1992,Siddappa1989,Nayak1993} \\ 
92 & U & 22.16-1613 & 0.5-160 & 262 & \cite{Barros1980,Muckenheim1980,Hardie1970,Hardie1971,Bui1989,Goldzahl1957,Goncalves1986,Kahane1989,Kasten1986,Lestone1988,Hauser1966,Shahi1998,Shahi1997,Kumar2009,Kumar2001,Nayak1992,Siddappa1989,Nayak1993,Taylor1987} \\ 
\hline 
\end{tabular}
\end{center}
\end{table*}

Some experimental cross sections have been published only in graphical form;
numerical values were digitized from the plots by means of the Engauge
\cite{engauge} software.
The error introduced by the digitization process was estimated by digitizing a
few plots, whose numerical content is explicitly reported in the related
publications.
The reliability of the digitized cross section values is hindered by the
difficulty of appraising the experimental points and their error bars in plots that
may span several orders of magnitude in logarithmic scale; these digitized data
were not used in the validation analysis, if found incompatible with other
measurements in the same experimental configuration (target element, incident
photon energy and measured angle) reported in the literature in numerical form.



Large discrepancies are evident in some of the experimental data;
systematic effects are likely present in some cases, where the Wald-Wolfowitz
test \cite{wald} detects sequences of positive or negative differences between
data samples originating from different experimental groups, which are 
incompatible with randomness.
Experimental data exhibiting large discrepancies with respect to other
measurements in similar configurations, that hint to the presence of systematic
effects, are excluded from the validation process.

Correct estimate of experimental errors is a concern in the validation process,
since unrealistic estimation of the experimental errors may lead to incorrect
conclusions regarding the rejection of the null hypothesis in tests whose statistic
takes into account experimental uncertainties explicitly.
Although technological developments have contributed to improved precision of
measurement, it has been remarked that in some circumstances estimates of
experimental uncertainties of the order of 5\% or less reported in the
literature may be optimistic \cite{bradley1999}.
Experimental measurements claiming much smaller uncertainties than similar ones
have been critically evaluated in the analysis process; effects related to the
validation of simulation models are discussed in section \ref{sec_diffcs}.



\subsection{Data analysis method}
\label{sec_method}

The evaluation of the simulation models performed in this study has two
objectives: to validate them quantitatively, and to compare their relative capabilities.

The scope of the software validation process is defined according to the
guidelines of the pertinent IEEE Standard \cite{ieee_vv}, which conforms to the
life cycle process standard defined in ISO/IEC Standard 12207 \cite{iso12207}.
For the problem domain considered in this paper, the validation process provides
evidence that the software models photon-atom elastic scattering consistent with
experiment.

A quantitative analysis, based on statistical methods, is practically possible
only for the validation of differential cross sections, for which a large sample
of experimental data is available.
The analysis of differential cross sections is articulated over two stages: the
first one estimates the compatibility between the values calculated by
each simulation model and experimental data, while the second exploits the
results of the first stage to determine whether the various models exhibit any
significant differences in their compatibility with experiment.

The first stage encompasses a number of test cases, each one
corresponding to a photon energy for which experimental data are available.
For each test case, cross sections calculated by the software are compared with
measured ones by means of goodness-of-fit tests; the null hypothesis in the
tests is defined as the equivalence of the simulated and experimental data
distributions subject to comparison.
The Statistical Toolkit \cite{gof1,gof2} is exploited for these tests.
The level of significance of the tests described in the following sections is
0.01, unless differently stated; it means that the null hypothesis is rejected
whenever the p-value resulting from the test statistic is smaller than 0.01.

The goodness-of-fit analysis is primarily based on the $\chi^2$ test \cite{bock}.
Among goodness-of-fit tests, this test has the peculiarity of taking into account
experimental uncertainties explicitly; therefore the test statistic is sensitive
to their correct appraisal.
Three tests based on the empirical distribution function, the
Kolmogorov-Smirnov\cite{kolmogorov1933,smirnov1939}, Anderson-Darling
\cite{anderson1952,anderson1954} and Cramer-von Mises
\cite{cramer1928,vonmises1931} tests, have also been applied;
nevertheless, as documented in the following sections, they exhibit limited
discriminant power over the data sample of this study.
In this context one may want to recall that the power of goodness-of-fit
tests is still object of active research in statistics.

The ``efficiency'' of a physics model is defined as the fraction of test cases
in which the $\chi^2$ test does not reject the null hypothesis at 0.01
level of significance: it quantifies the capability of that simulation model to
produce results statistically consistent with experiment over the whole set of
test cases, which in physical terms means over the whole energy range of the 
experimental data sample involved in the validation process.

The second stage of the statistical analysis quantifies the differences of the
simulation models in compatibility with experiment.
It consists of a categorical analysis based on contingency tables, which derive
from the results of the $\chi^2$ test: the outcome of this test is classified as
``fail'' or ``pass'', according respectively to whether the hypothesis of
compatibility of experimental and calculated data is rejected or not.
The categorical analysis takes as a reference the simulation model exhibiting
the largest efficiency at reproducing experimental data; the other models are
compared to it, to determine whether they exhibit statistically significant
differences of compatibility with measurements.

The null hypothesis in the analysis of a contingency table assumes the
equivalent compatibility with experiment of the models it compares.
Contingency tables are analyzed with Fisher's exact test \cite{fisher} and
with the $\chi^2$ test applying Yates' continuity correction \cite{yates}; the
latter ensures meaningful results when one or more cells of the table
have few entries.
Pearson's $\chi^2$ test \cite{pearson} is also performed on contingency tables,
when their content is consistent with its applicability.
The use of different tests in the analysis of contingency tables contributes to
the robustness of the results, as it mitigates the risk of introducing
systematic effects, which could be due to the peculiar mathematical features of
a single test.
 
The significance level for the rejection of the null hypothesis in the analysis
of contingency tables is 0.05, unless differently specified.

The scarcity of experimental measurements of elastic scattering total cross
sections in the literature hinders the validation of simulation models through
similar statistical analysis methods: only qualitative general considerations
can be made, while quantitative assessments are limited by the small sample of
published experimental data.
Nevertheless, the results of the analysis of differential cross sections 
contribute indirectly to assess the validity of total cross
sections, since some of the implemented total cross section models derive from
the integration of the differential cross sections subject to validation.

\begin{figure*}
\centerline{\includegraphics[angle=0,width=18cm]{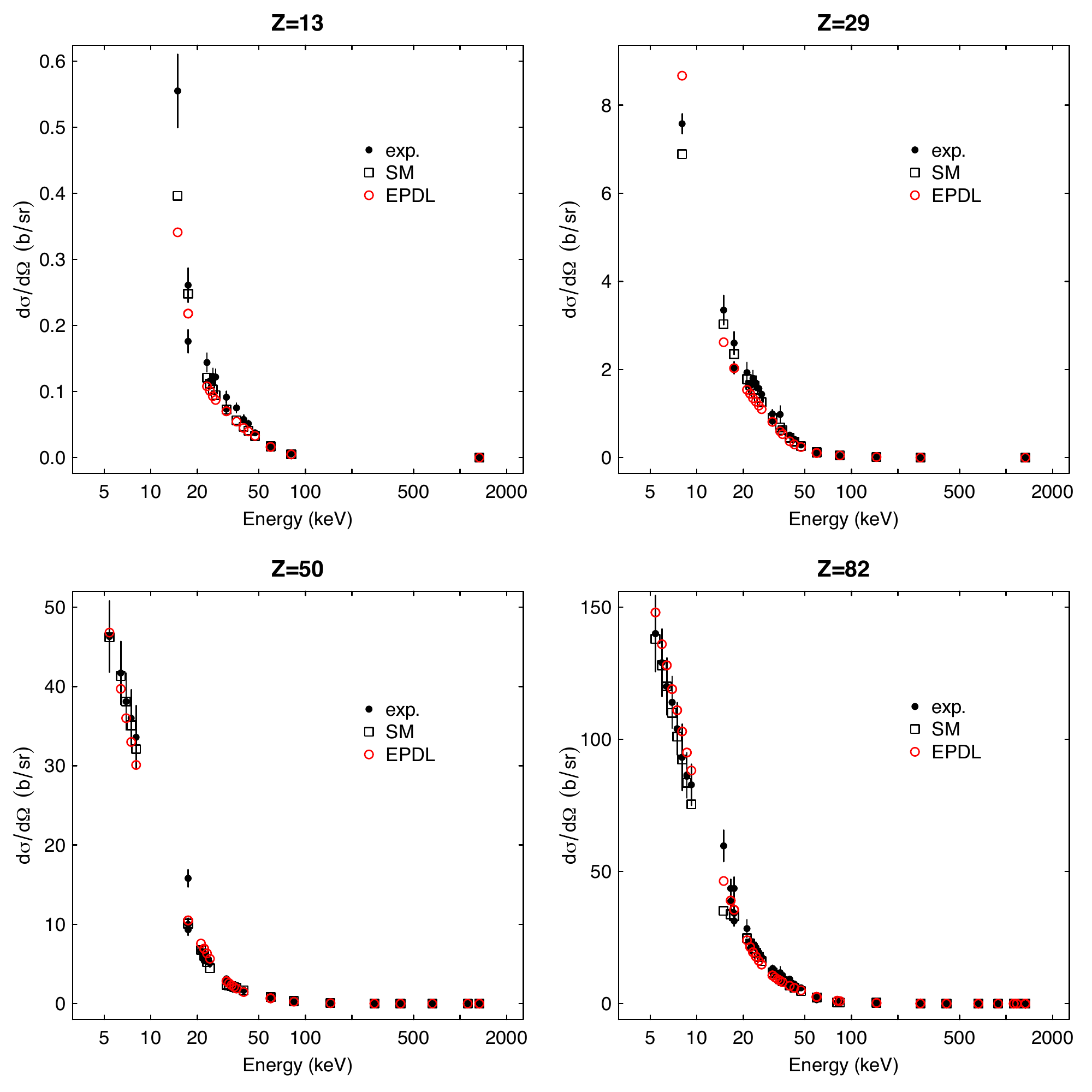}}
\caption{Differential cross section as a function of energy at
$\theta=90^{\circ}$ for aluminium, copper, tin and lead: experimental
measurements (black  circles), calculations based on S-matrix (SM, black empty
squares) and on EPDL (red  circles).
The sources of experimental data are documented in Tables \ref{tab_exp1}
and \ref{tab_exp2}.}
\label{fig_plot90}
\end{figure*}

\begin{figure*}
\centerline{\includegraphics[angle=0,width=18cm]{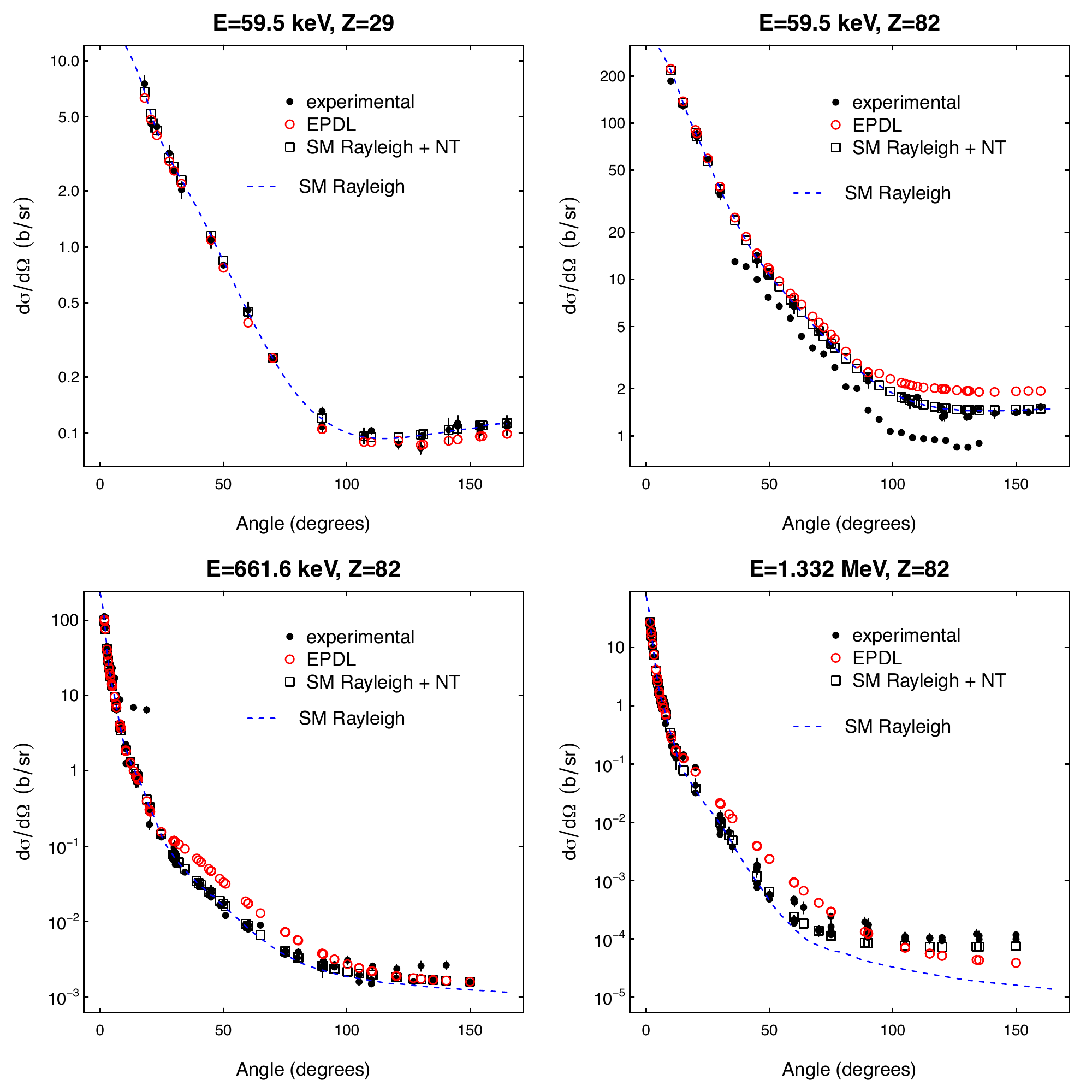}}
\caption{Differential cross section as a function of scattering angle for representative
energies and target elements: experimental measurements (black circles),
calculations based on S-matrix (SM, black empty squares) and on EPDL (red
circles). The S-matrix calculations account for Rayleigh scattering and nuclear
Thomson scattering; S-matrix calculations limited to the Rayleigh scattering
amplitude are shown as a blue dashed line.
The sources of experimental data are documented in Tables \ref{tab_exp1}
and \ref{tab_exp2}.}
\label{fig_plot_theta}
\end{figure*}

\begin{figure*}
\centerline{\includegraphics[angle=0,width=18cm]{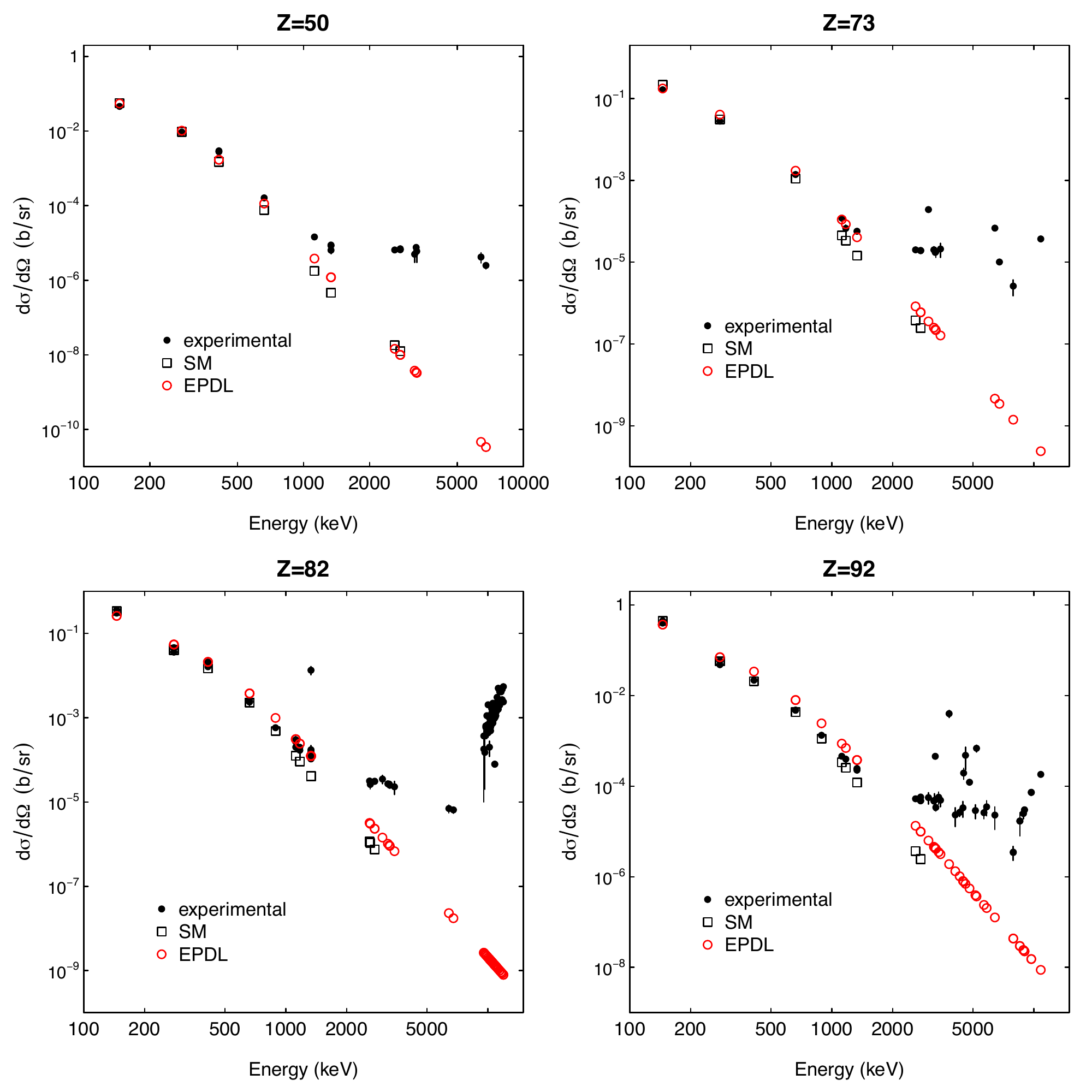}}
\caption{Differential cross section as a function of energy at
$\theta=90^{\circ}$ for tin, tantalum, lead and uranium: experimental
measurements (black circles), calculations based on S-matrix (SM, black empty
squares) and on EPDL (red circles).
The sources of experimental data are documented in Tables \ref{tab_exp1}
and \ref{tab_exp2}.}
\label{fig_he}
\end{figure*}


\section{Results}
\label{sec_results}

The following sections report the results of the comparison of differential and total
cross sections  with experimental data, and of the evaluation of the computational
performance of the various simulation models.



\subsection{General features}
\label{sec_general}


Elastic scattering cross sections vary by several orders of magnitude over 
the data sample involved in the validation analysis.
Cross sections decrease as a function of photon energy and scattering angle, and
are larger for heavy elements: these trends are illustrated in
Figs.~\ref{fig_plot90} and \ref{fig_plot_theta}, that show experimental photon
elastic scattering measurements along with the values calculated by
representative simulation models for the same energy and scattering angle
settings.

Experimental differential cross sections are the result of all the physics processes that
contribute to photon elastic scattering, while the simulation models evaluated
in this paper account for the Rayleigh scattering amplitude only or, in the case
of the model based on S-matrix calculations, for the sum of Rayleigh and Thomson
scattering amplitudes.
This feature is evident in Fig. \ref{fig_he}, which includes some of the
higher energy measurements in the experimental data sample: other processes,
such as Delbr\"uck scattering, should be taken into account in the simulation,
along with Rayleigh scattering, to model photon elastic scattering accurately at
higher energies.
The plots also expose some characteristics of the experimental data: systematic
effects affecting some of the measurements, and the presence of outliers in the
experimental sample.


%
%

\begin{figure}
\centerline{\includegraphics[angle=0,width=9cm]{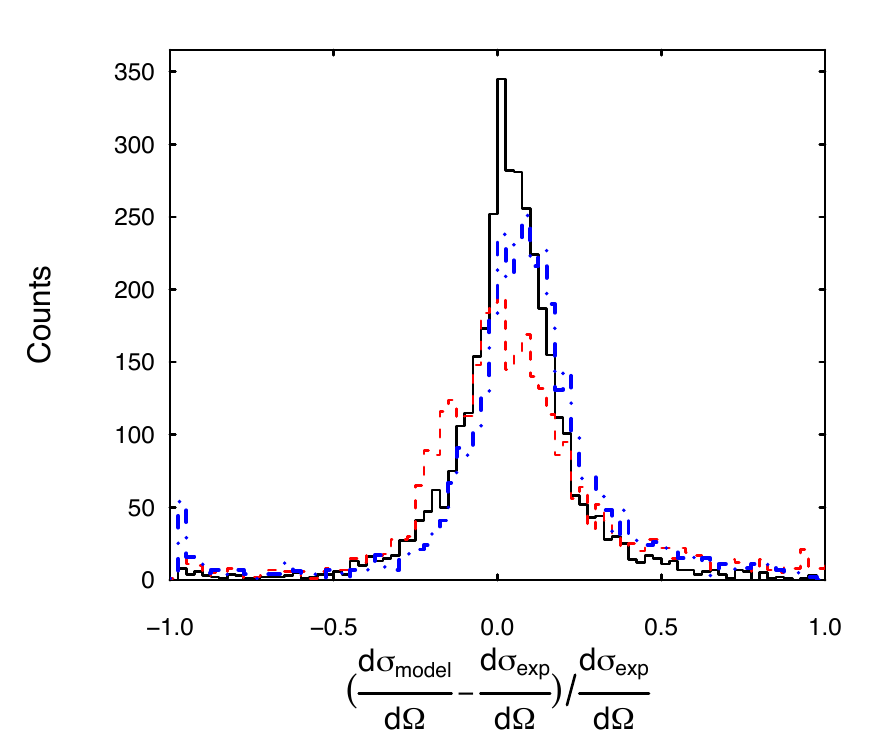}}
\caption{Relative difference between the differential cross sections calculated
by simulation models and the corresponding experimental measurements:
calculations based on based S-matrix (SM, solid black line), EPDL (dashed red line)
and MFASF form factors (dot-dashed blue line). 
The plot is limited to photon energies up to 1.5 MeV }
\label{fig_diff}
\end{figure}

\begin{figure}
\centerline{\includegraphics[angle=0,width=9cm]{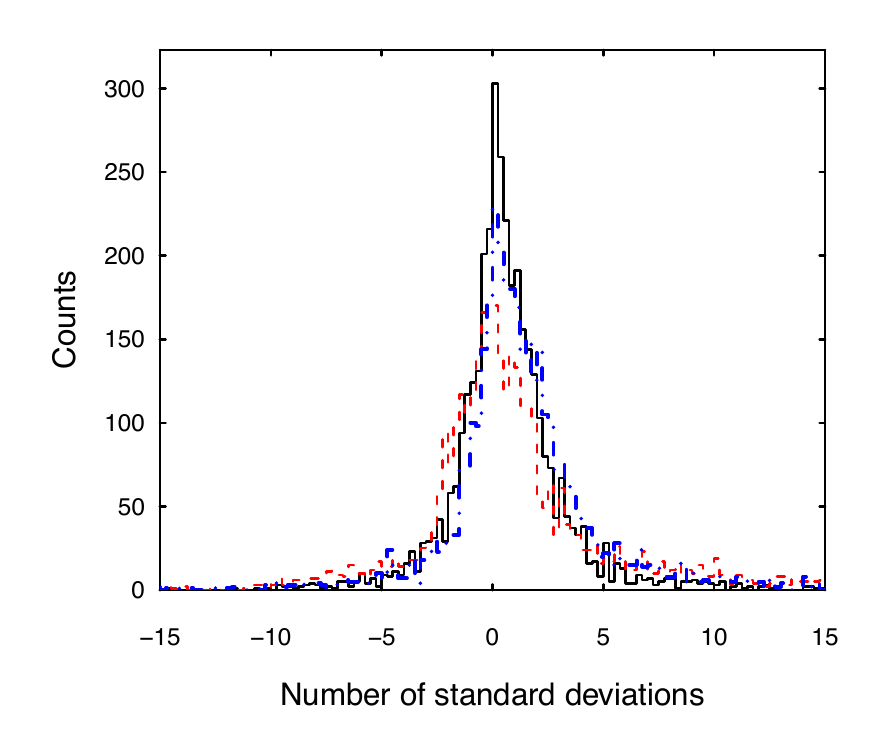}}
\caption{Difference between the differential cross sections calculated by
simulation models and the corresponding experimental measurements, expressed in
terms of number of standard deviations (experimental uncertainties):
calculations based on based S-matrix (SM, solid black line), EPDL (dashed red line)
and MFASF form factors (dot-dashed blue line).
The plot is limited to photon energies up to 1.5 MeV }
\label{fig_nsigma}
\end{figure}

The distributions in Figs. \ref{fig_diff} and \ref{fig_nsigma} illustrate the
difference between calculated and measured differential cross sections, for a
few representative models: two options based on the form factor approximation,
respectively using the form factors tabulated in EPDL97 and modified form factors
with anomalous scattering factors, and the model based on S-matrix calculations.
Fig. \ref{fig_diff} shows the relative difference between simulated and
experimental values.
Fig. \ref{fig_nsigma} accounts for the different precision of the data in the
experimental sample: it represents the difference between calculated and
experimental differential cross sections in units of standard deviations, i.e.
of the uncertainties associated with the experimental measurements.
These plots are limited to photon energies up to 1.5 MeV, since, as shown in
Fig. \ref{fig_he}, at higher energies other processes 
should be taken into account to describe photon elastic scattering.
The long tails of the distributions in Figs. \ref{fig_diff} and \ref{fig_nsigma}
are the result of multiple factors: neglect of amplitudes other than Rayleigh
scattering in some of the calculations, inadequacy of the simulation models, 
and contamination of the experimental
sample by measurements affected by systematic effects and outliers.
One can observe in Fig. \ref{fig_diff} that the distribution of the relative differences
between the differential cross sections calculated by the models and 
experimental data is narrower for the S-matrix model, while it is wider for the
cross sections based on EPDL and on modified form factors with anomalous
scattering factor corrections.

\begin{figure}
\centerline{\includegraphics[angle=0,width=9cm]{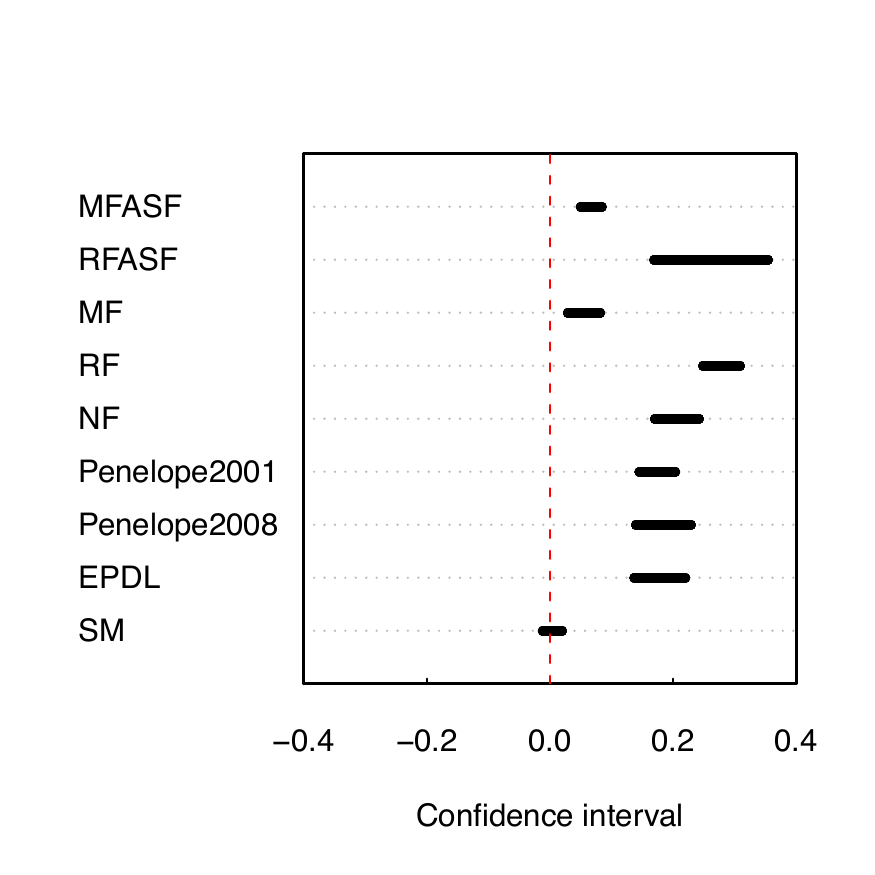}}
\caption{Confidence intervals for the true mean difference between differential
cross sections calculated by Rayleigh scattering simulation models and
experimental measurements; the dashed red line identifies the value of zero in
the dot chart.}
\label{fig_tconf}
\end{figure}

The one sample t-test \cite{barlow} for the hypothesis of null mean difference between
simulation and experiment confirms the qualitative features observed in 
Figs. \ref{fig_diff} and \ref{fig_nsigma}.
Fig. \ref{fig_tconf} shows the 99\% confidence intervals for the true mean difference
associated with the various models: the model based on S-matrix calculations is
the only one for which the confidence interval includes zero.
The hypothesis of null mean difference from experiment 
equal to zero is rejected for all the cross section
models with 0.001 significance, with the exception of the model based on
S-matrix calculations, for which the p-value resulting from the t-test is 0.500.

\begin{figure}
\centerline{\includegraphics[angle=0,width=9cm]{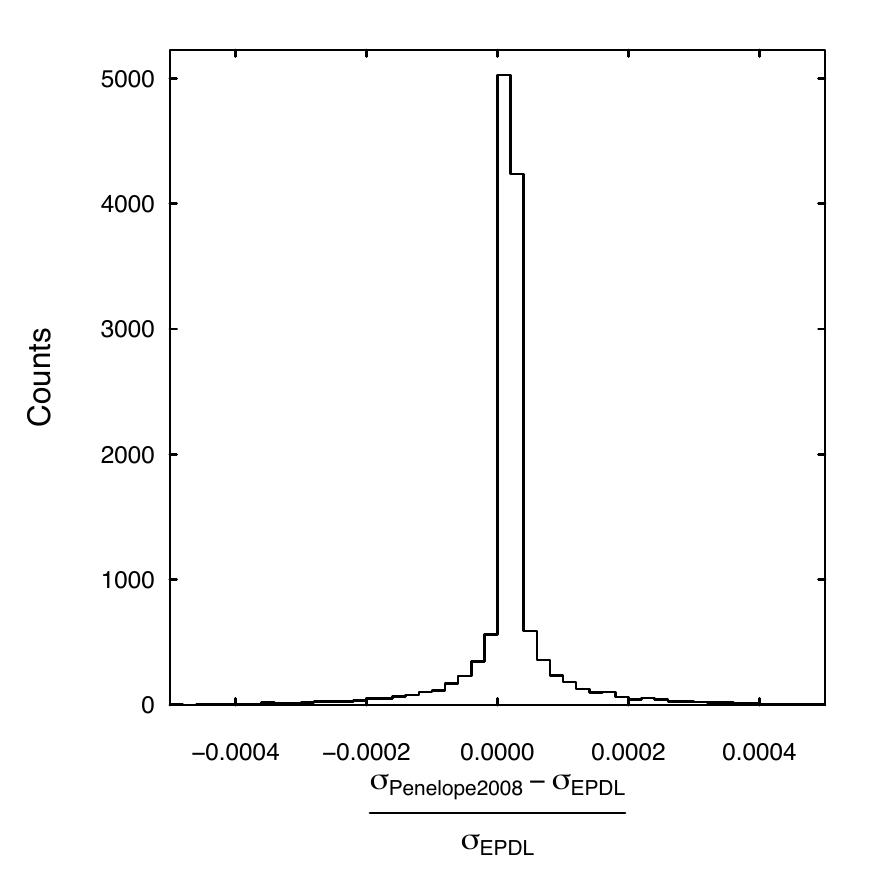}}
\caption{Relative difference between the total cross sections based on
interpolation of Penelope 2008-2011 tabulations derived from EPDL97 and by
interpolation of native EPDL97 tabulations as described in \cite{lowe_e}.}
\label{fig_penelopetot}
\end{figure}

\begin{figure}
\centerline{\includegraphics[angle=0,width=9cm]{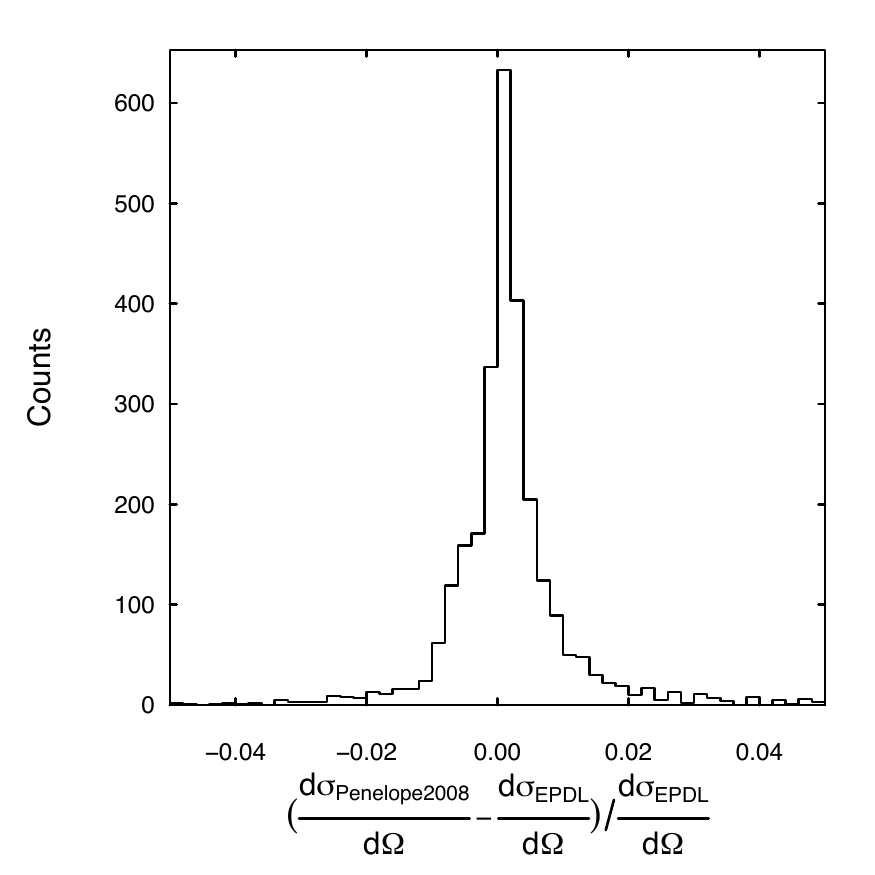}}
\caption{Relative difference between the differential cross sections calculated
based on interpolation of Penelope 2008-2011 tabulated form factors derived from
EPDL97 and by interpolation of native EPDL97 tabulations, as described in
\cite{lowe_e}. The distribution has mean equal to -0.0007 and standard deviation
equal to 0.01.}
\label{fig_penelope}
\end{figure}

The algorithm for Rayleigh scattering simulation reengineered from Penelope
2008-2011 versions produces very similar results to the algorithm originally
developed in \cite{lowe_e}, which corresponds to the policy class identified in
this paper as EPDL: in fact, both of them are based on EPDL97 data.
Figs.~\ref{fig_penelopetot} and \ref{fig_penelope} show the relative difference
between the total and differential cross sections calculated by these two
models; the small differences are due to the different reference grids used
by the two models in the respective tabulations of total cross sections and
non-relativistic form factors, which affect the values calculated by interpolation.
The difference of resulting total cross sections appears insignificant,
considering that the scale of Fig. \ref{fig_penelopetot} extends to relative
differences up to 0.05\%, while the distribution of differences between differential cross sections,
shown in Fig. \ref{fig_penelope}, is wider, having
a standard deviation of 1\%: the effect of these differences on the accuracy of
the two models is quantitatively estimated in section \ref{sec_diffcs}.

\subsection{Differential cross sections}
\label{sec_diffcs}

The analysis of differential cross sections is articulated over the set of
discrete energies for which experimental data are available.
The p-values resulting from the $\chi^{2}$ test applied to the data samples
corresponding to each energy are listed in Tables \ref{tab_pvalue_all},
\ref{tab_pvalue_90} and \ref{tab_pvalue_gt90}, respectively concerning the whole
angular range, data up to 90$^{\circ}$ and above 90$^{\circ}$ (backscattering).


\begin{table*}[htbp]
  \centering
  \caption{Differential corss sections: p-values resulting from the $\chi^{2}$ test over the whole angular range}
    \begin{tabular}{c|rrrrrrrrrr}
    \hline
    \textbf{E } & {\textbf{Penelope}} & {\textbf{Penelope}} & {\textbf{EPDL}} & {\textbf{EPDL}} & {\textbf{SM}} & {\textbf{RF}} & {\textbf{NF}} & {\textbf{MF}} & {\textbf{MF}} & {\textbf{RF}} \\
    (keV) & {\textbf{2001}} & {\textbf{2008-2011}} & {\textbf{}} & {\textbf{ASF}} & {} & {\textbf{}} & {\textbf{}} & {\textbf{}} & {\textbf{ASF}} & {\textbf{ASF}} \\
    \hline
    5.45  & 0.998 & 0.996 & 0.992 & 0.297 & 1.000 & 0.952 & 0.996 & 0.999 & 1.000 & 1.000 \\
    5.93  & 0.706 & 0.745 & 0.741 & 0.049 & 0.999 & 0.432 & 0.732 & 0.775 & 1.000 & 1.000 \\
    6.4   & 0.860 & 0.830 & 0.863 & 0.148 & 1.000 & 0.728 & 0.829 & 0.923 & 1.000 & 1.000 \\
    6.93  & 0.913 & 0.877 & 0.876 & 0.008 & 1.000 & 0.722 & 0.883 & 0.944 & 1.000 & 1.000 \\
    7.47  & 0.695 & 0.626 & 0.614 & 0.041 & 1.000 & 0.386 & 0.637 & 0.749 & 1.000 & 1.000 \\
    8.04  & 0.053 & 0.001 & 0.001 & $<0.001$ & 0.696 & $<0.001$ & 0.001 & $<0.001$ & 0.736 & 0.727 \\
    8.63  & 0.144 & 0.249 & 0.241 & $<0.001$ & 0.759 & 0.020 & 0.199 & 0.232 & 0.820 & 0.812 \\
    9.23  & 0.023 & 0.033 & 0.030 & $<0.001$ & 0.652 & 0.002 & 0.024 & 0.029 & 0.719 & 0.708 \\
    13.95 & $<0.001$ & $<0.001$ & $<0.001$ & 0.002 & 0.154 & $<0.001$ & $<0.001$ & $<0.001$ & 0.046 & 0.064 \\
    14.93 & $<0.001$ & $<0.001$ & $<0.001$ & $<0.001$ & $<0.001$ & 0.001 & $<0.001$ & $<0.001$ & $<0.001$ & $<0.001$ \\
    16.58 & 0.956 & 0.908 & 0.901 & 0.587 & 0.218 & 1.000 & 0.947 & 0.947 & 0.305 & 0.281 \\
    17.44 & $<0.001$ & $<0.001$ & $<0.001$ & $<0.001$ & 0.102 & $<0.001$ & $<0.001$ & $<0.001$ & 0.013 & 0.025 \\
    17.75 & $<0.001$ & $<0.001$ & $<0.001$ & $<0.001$ & 0.105 & $<0.001$ & $<0.001$ & $<0.001$ & 0.009 & 0.025 \\
    21.12 & 0.008 & 0.001 & 0.001 & 0.017 & 0.099 & $<0.001$ & 0.001 & 0.001 & 0.124 & 0.118 \\
    22.1  & $<0.001$ & $<0.001$ & $<0.001$ & $<0.001$ & 0.014 & $<0.001$ & $<0.001$ & $<0.001$ & $<0.001$ & $<0.001$ \\
    23.1  & $<0.001$ & $<0.001$ & $<0.001$ & $<0.001$ & $<0.001$ & 0.002 & $<0.001$ & $<0.001$ & $<0.001$ & $<0.001$ \\
    24.14 & $<0.001$ & $<0.001$ & $<0.001$ & $<0.001$ & $<0.001$ & 0.002 & $<0.001$ & $<0.001$ & $<0.001$ & $<0.001$ \\
    25.2  & $<0.001$ & $<0.001$ & $<0.001$ & $<0.001$ & $<0.001$ & $<0.001$ & $<0.001$ & $<0.001$ & $<0.001$ & $<0.001$ \\
    26.27 & $<0.001$ & $<0.001$ & $<0.001$ & $<0.001$ & $<0.001$ & 0.001 & $<0.001$ & $<0.001$ & 0.001 & $<0.001$ \\
    26.36 & $<0.001$ & $<0.001$ & $<0.001$ & $<0.001$ & $<0.001$ & $<0.001$ & $<0.001$ & $<0.001$ & 0.772 & 0.758 \\
    30.85 & $<0.001$ & 0.280 & 0.322 & $<0.001$ & 0.011 & 0.955 & 0.399 & 0.507 & 0.028 & 0.015 \\
    30.97 & 0.001 & 0.011 & 0.020 & 0.004 & 0.377 & 0.001 & 0.006 & 0.002 & 0.667 & 0.509 \\
    32.06 & 0.003 & 0.078 & 0.100 & 0.197 & 0.602 & 0.030 & 0.066 & 0.036 & 0.885 & 0.749 \\
    33.3  & 0.004 & 0.077 & 0.082 & 0.048 & 0.306 & 0.154 & 0.104 & 0.072 & 0.589 & 0.422 \\
    34.6  & $<0.001$ & $<0.001$ & $<0.001$ & 0.005 & 0.033 & 0.006 & 0.001 & 0.001 & 0.071 & 0.046 \\
    35.86 & 0.001 & 0.001 & 0.001 & 0.008 & 0.148 & 0.195 & 0.002 & 0.002 & 0.314 & 0.179 \\
    39.9  & $<0.001$ & $<0.001$ & $<0.001$ & $<0.001$ & 0.061 & 0.043 & 0.004 & 0.003 & 0.083 & 0.072 \\
    42.75 & $<0.001$ & 0.001 & 0.001 & 0.002 & 0.183 & 0.001 & 0.003 & 0.002 & 0.082 & 0.089 \\
    47.24 & $<0.001$ & 0.011 & 0.021 & 0.003 & 0.019 & 0.005 & 0.002 & 0.065 & 0.022 & 0.044 \\
    59.54 & $<0.001$ & $<0.001$ & $<0.001$ & $<0.001$ & $<0.001$ & $<0.001$ & $<0.001$ & $<0.001$ & $<0.001$ & $<0.001$ \\
    81    & $<0.001$ & $<0.001$ & $<0.001$ & $<0.001$ & $<0.001$ & $<0.001$ & $<0.001$ & $<0.001$ & $<0.001$ & $<0.001$ \\
    84.3  & $<0.001$ & $<0.001$ & $<0.001$ & $<0.001$ & $<0.001$ & $<0.001$ & $<0.001$ & $<0.001$ & $<0.001$ & $<0.001$ \\
    88    & $<0.001$ & $<0.001$ & $<0.001$ & $<0.001$ & $<0.001$ & $<0.001$ & $<0.001$ & $<0.001$ & $<0.001$ & $<0.001$ \\
    122   & 0.196 & 0.500 & 0.541 & $<0.001$ & 0.107 & $<0.001$ & 0.478 & 0.287 & 0.033 & 0.182 \\
    136   & 0.534 & 0.837 & 0.790 & $<0.001$ & 0.913 & 0.047 & 0.893 & 0.836 & 0.669 & 0.891 \\
    145.4 & $<0.001$ & $<0.001$ & $<0.001$ & $<0.001$ & 0.037 & $<0.001$ & $<0.001$ & $<0.001$ & $<0.001$ & $<0.001$ \\
    244.7 & $<0.001$ & 0.025 & 0.032 & 0.006 & 0.038 & $<0.001$ & 0.027 & 0.024 & 0.037 & 0.036 \\
    279.2 & $<0.001$ & $<0.001$ & $<0.001$ & $<0.001$ & 0.201 & $<0.001$ & $<0.001$ & $<0.001$ & $<0.001$ & $<0.001$ \\
    317   & $<0.001$ & $<0.001$ & $<0.001$ & 0.019 & $<0.001$ & 0.200 & $<0.001$ & $<0.001$ & $<0.001$ & $<0.001$ \\
    320   & $<0.001$ & $<0.001$ & $<0.001$ & $<0.001$ & $<0.001$ & $<0.001$ & $<0.001$ & $<0.001$ & $<0.001$ & $<0.001$ \\
    334   & 0.031 & 0.873 & 0.801 & $<0.001$ & 0.118 & $<0.001$ & 0.617 & 0.757 & $<0.001$ & 0.022 \\
    344   & $<0.001$ & 0.047 & 0.033 & $<0.001$ & 0.080 & $<0.001$ & 0.040 & 0.140 & $<0.001$ & $<0.001$ \\
    411.8 & $<0.001$ & $<0.001$ & $<0.001$ & $<0.001$ & 0.136 & $<0.001$ & $<0.001$ & $<0.001$ & $<0.001$ & $<0.001$ \\
    443.96 & $<0.001$ & $<0.001$ & $<0.001$ & $<0.001$ & 0.053 & $<0.001$ & $<0.001$ & 0.111 & 0.012 & $<0.001$ \\
    465   & 0.758 & 0.542 & 0.518 & 0.488 & 0.285 & 0.120 & 0.535 & 0.302 & 0.287 & 0.290 \\
    468.1 & $<0.001$ & $<0.001$ & $<0.001$ & $<0.001$ & 0.053 & $<0.001$ & $<0.001$ & $<0.001$ & $<0.001$ & $<0.001$ \\
    511   & 0.754 & 0.752 & 0.756 & 0.756 & 0.730 & 0.746 & 0.753 & 0.727 & 0.731 & 0.731 \\
    661.6 & $<0.001$ & $<0.001$ & $<0.001$ & $<0.001$ & 0.044 & $<0.001$ & $<0.001$ & $<0.001$ & $<0.001$ & $<0.001$ \\
    723   & $<0.001$ & $<0.001$ & $<0.001$ & $<0.001$ & 0.096 & $<0.001$ & $<0.001$ & 0.114 & 0.009 & $<0.001$ \\
    779   & $<0.001$ & $<0.001$ & $<0.001$ & $<0.001$ & 0.683 & $<0.001$ & $<0.001$ & 0.333 & $<0.001$ & $<0.001$ \\
    811   & 4.561 & 6.060 & 6.245 & 6.295 & 5.822 & 8.857 & 6.174 & 5.828 & 5.881 & 5.784 \\
    847   & $<0.001$ & $<0.001$ & $<0.001$ & $<0.001$ & 0.598 & $<0.001$ & $<0.001$ & 0.870 & 0.278 & $<0.001$ \\
    868   & 0.005 & 0.162 & 0.141 & 0.043 & 0.919 & $<0.001$ & 0.142 & 0.925 & 0.854 & $<0.001$ \\
    878   & 0.026 & 0.140 & 0.119 & 0.115 & 0.164 & 0.001 & 0.123 & 0.153 & 0.140 & 0.156 \\
    889   & $<0.001$ & $<0.001$ & $<0.001$ & $<0.001$ & 0.323 & $<0.001$ & $<0.001$ & $<0.001$ & $<0.001$ & $<0.001$ \\
    952   & $<0.001$ & $<0.001$ & $<0.001$ & $<0.001$ & $<0.001$ & $<0.001$ & $<0.001$ & $<0.001$ & $<0.001$ & $<0.001$ \\
    964   & $<0.001$ & $<0.001$ & $<0.001$ & $<0.001$ & 0.152 & $<0.001$ & $<0.001$ & $<0.001$ & $<0.001$ & $<0.001$ \\
    1005  & $<0.001$ & $<0.001$ & $<0.001$ & $<0.001$ & 0.987 & $<0.001$ & $<0.001$ & 0.968 & 0.697 & $<0.001$ \\
    1019  & 0.280 & 0.053 & 0.045 & 0.045 & 0.075 & $<0.001$ & 0.049 & 0.073 & 0.066 & 0.095 \\
    1038  & $<0.001$ & 0.041 & 0.011 & 0.011 & 0.004 & $<0.001$ & 0.056 & 0.019 & $<0.001$ & $<0.001$ \\
    1086  & $<0.001$ & $<0.001$ & $<0.001$ & $<0.001$ & 0.125 & $<0.001$ & $<0.001$ & 0.011 & $<0.001$ & $<0.001$ \\
    1112  & $<0.001$ & $<0.001$ & $<0.001$ & $<0.001$ & 0.060 & $<0.001$ & $<0.001$ & $<0.001$ & $<0.001$ & $<0.001$ \\
    1120  & $<0.001$ & $<0.001$ & $<0.001$ & $<0.001$ & 0.001 & $<0.001$ & $<0.001$ & $<0.001$ & $<0.001$ & $<0.001$ \\
    1173  & $<0.001$ & $<0.001$ & $<0.001$ & $<0.001$ & 0.029 & $<0.001$ & $<0.001$ & $<0.001$ & $<0.001$ & $<0.001$ \\
    1189  & 0.948 & 0.401 & 0.380 & 0.380 & 0.523 & 0.022 & 0.384 & 0.482 & 0.468 & 0.522 \\
    1238  & $<0.001$ & $<0.001$ & $<0.001$ & $<0.001$ & 0.011 & $<0.001$ & $<0.001$ & 0.008 & $<0.001$ & $<0.001$ \\
    1274  & $<0.001$ & $<0.001$ & $<0.001$ & $<0.001$ & 0.076 & $<0.001$ & $<0.001$ & 0.016 & $<0.001$ & $<0.001$ \\
    1302  & 0.054 & 0.039 & 0.031 & 0.031 & 0.050 & $<0.001$ & 0.032 & 0.048 & 0.044 & 0.057 \\
    1332  & $<0.001$ & $<0.001$ & $<0.001$ & $<0.001$ & $<0.001$ & $<0.001$ & $<0.001$ & $<0.001$ & $<0.001$ & $<0.001$ \\
    1359  & $<0.001$ & $<0.001$ & $<0.001$ & $<0.001$ & 0.382 & $<0.001$ & $<0.001$ & 0.325 & 0.133 & $<0.001$ \\
    1408  & $<0.001$ & $<0.001$ & $<0.001$ & $<0.001$ & 0.312 & $<0.001$ & $<0.001$ & 0.068 & $<0.001$ & $<0.001$ \\
    $\geq$1600 & $<0.001$ & $<0.001$ & $<0.001$ & $<0.001$ & $<0.001$ & $<0.001$ & $<0.001$ & $<0.001$ & $<0.001$ & $<0.001$ \\
    \end{tabular}%
  \label{tab_pvalue_all}%
\end{table*}%

\begin{table*}[htbp]
  \centering
  \caption{Differential cross sections: p-values resulting from the $\chi^{2}$ test, data sample up to 90$^{\circ}$}
    \begin{tabular}{c|rrrrrrrrrr}
    \hline
    \textbf{E } & {\textbf{Penelope}} & {\textbf{Penelope}} & {\textbf{EPDL}} & {\textbf{EPDL}} & {\textbf{SM}} & {\textbf{RF}} & {\textbf{NF}} & {\textbf{MF}} & {\textbf{MF}} & {\textbf{RF}} \\
    (keV) & {\textbf{2001}} & {\textbf{2008-2011}} & {\textbf{}} & {\textbf{ASF}} & {} & {\textbf{}} & {\textbf{}} & {\textbf{}} & {\textbf{ASF}} & {\textbf{ASF}} \\
    \hline
    5.45  & 0.998 & 0.996 & 0.992 & 0.297 & 1.000 & 0.952 & 0.996 & 0.999 & 1.000 & 1.000 \\
    5.9   & 0.706 & 0.745 & 0.741 & 0.049 & 0.999 & 0.432 & 0.732 & 0.775 & 1.000 & 1.000 \\
    6.40  & 0.860 & 0.830 & 0.863 & 0.148 & 1.000 & 0.728 & 0.829 & 0.923 & 1.000 & 1.000 \\
    6.93  & 0.913 & 0.877 & 0.876 & 0.008 & 1.000 & 0.722 & 0.883 & 0.944 & 1.000 & 1.000 \\
    7.47  & 0.695 & 0.626 & 0.614 & 0.041 & 1.000 & 0.386 & 0.637 & 0.749 & 1.000 & 1.000 \\
    8.04  & 0.053 & 0.001 & 0.001 & $<0.001$ & 0.696 & $<0.001$ & 0.001 & $<0.001$ & 0.736 & 0.727 \\
    8.63  & 0.144 & 0.249 & 0.241 & $<0.001$ & 0.759 & 0.020 & 0.199 & 0.232 & 0.820 & 0.812 \\
    9.23  & 0.023 & 0.033 & 0.030 & $<0.001$ & 0.652 & 0.002 & 0.024 & 0.029 & 0.719 & 0.708 \\
    14.93 & $<0.001$ & $<0.001$ & $<0.001$ & $<0.001$ & $<0.001$ & 0.001 & $<0.001$ & $<0.001$ & $<0.001$ & $<0.001$ \\
    16.58 & 0.956 & 0.908 & 0.901 & 0.587 & 0.218 & 1.000 & 0.947 & 0.947 & 0.305 & 0.281 \\
    17.4  & $<0.001$ & $<0.001$ & $<0.001$ & $<0.001$ & 0.102 & $<0.001$ & $<0.001$ & $<0.001$ & 0.013 & 0.025 \\
    21.12 & 0.008 & 0.001 & 0.001 & 0.017 & 0.099 & $<0.001$ & 0.001 & 0.001 & 0.124 & 0.118 \\
    22.1  & $<0.001$ & $<0.001$ & $<0.001$ & $<0.001$ & 0.159 & $<0.001$ & $<0.001$ & $<0.001$ & 0.021 & 0.040 \\
    23.10 & $<0.001$ & $<0.001$ & $<0.001$ & $<0.001$ & $<0.001$ & 0.002 & $<0.001$ & $<0.001$ & $<0.001$ & $<0.001$ \\
    24.14 & $<0.001$ & $<0.001$ & $<0.001$ & $<0.001$ & $<0.001$ & 0.002 & $<0.001$ & $<0.001$ & $<0.001$ & $<0.001$ \\
    25.2  & $<0.001$ & $<0.001$ & $<0.001$ & $<0.001$ & $<0.001$ & $<0.001$ & $<0.001$ & $<0.001$ & $<0.001$ & $<0.001$ \\
    26.27 & $<0.001$ & $<0.001$ & $<0.001$ & $<0.001$ & $<0.001$ & 0.001 & $<0.001$ & $<0.001$ & 0.001 & $<0.001$ \\
    30.9  & $<0.001$ & 0.280 & 0.322 & $<0.001$ & 0.011 & 0.955 & 0.399 & 0.507 & 0.028 & 0.015 \\
    31.0  & 0.001 & 0.011 & 0.020 & 0.004 & 0.377 & 0.001 & 0.006 & 0.002 & 0.667 & 0.509 \\
    32.06 & 0.003 & 0.078 & 0.100 & 0.197 & 0.602 & 0.030 & 0.066 & 0.036 & 0.885 & 0.749 \\
    33.30 & 0.004 & 0.077 & 0.082 & 0.048 & 0.306 & 0.154 & 0.104 & 0.072 & 0.589 & 0.422 \\
    34.60 & $<0.001$ & $<0.001$ & $<0.001$ & 0.005 & 0.033 & 0.006 & 0.001 & 0.001 & 0.071 & 0.046 \\
    35.86 & 0.001 & 0.001 & 0.001 & 0.008 & 0.148 & 0.195 & 0.002 & 0.002 & 0.314 & 0.179 \\
    39.9  & $<0.001$ & $<0.001$ & $<0.001$ & $<0.001$ & 0.061 & 0.043 & 0.004 & 0.003 & 0.083 & 0.072 \\
    42.75 & $<0.001$ & 0.001 & 0.001 & 0.002 & 0.183 & 0.001 & 0.003 & 0.002 & 0.082 & 0.089 \\
    47.24 & $<0.001$ & 0.011 & 0.021 & 0.003 & 0.019 & 0.005 & 0.002 & 0.065 & 0.022 & 0.044 \\
    59.54 & $<0.001$ & $<0.001$ & $<0.001$ & $<0.001$ & $<0.001$ & $<0.001$ & $<0.001$ & $<0.001$ & $<0.001$ & 0.001 \\
    81    & $<0.001$ & $<0.001$ & $<0.001$ & $<0.001$ & $<0.001$ & $<0.001$ & $<0.001$ & $<0.001$ & $<0.001$ & $<0.001$ \\
    84.3  & $<0.001$ & $<0.001$ & $<0.001$ & $<0.001$ & $<0.001$ & $<0.001$ & $<0.001$ & $<0.001$ & $<0.001$ & $<0.001$ \\
    122   & 0.196 & 0.500 & 0.541 & $<0.001$ & 0.107 & $<0.001$ & 0.478 & 0.287 & 0.033 & 0.182 \\
    136   & 0.534 & 0.837 & 0.790 & $<0.001$ & 0.913 & 0.047 & 0.893 & 0.836 & 0.669 & 0.891 \\
    145.4 & $<0.001$ & $<0.001$ & $<0.001$ & $<0.001$ & 0.459 & $<0.001$ & $<0.001$ & $<0.001$ & $<0.001$ & $<0.001$ \\
    244.7 & $<0.001$ & 0.025 & 0.032 & 0.006 & 0.038 & $<0.001$ & 0.027 & 0.024 & 0.037 & 0.036 \\
    279.2 & $<0.001$ & $<0.001$ & $<0.001$ & $<0.001$ & 0.096 & $<0.001$ & $<0.001$ & $<0.001$ & $<0.001$ & $<0.001$ \\
    317   & $<0.001$ & $<0.001$ & $<0.001$ & 0.019 & $<0.001$ & 0.200 & $<0.001$ & $<0.001$ & $<0.001$ & $<0.001$ \\
    320   & $<0.001$ & $<0.001$ & $<0.001$ & $<0.001$ & 1.000 & $<0.001$ & $<0.001$ & $<0.001$ & $<0.001$ & $<0.001$ \\
    334   & 0.031 & 0.873 & 0.801 & $<0.001$ & 0.118 & $<0.001$ & 0.617 & 0.757 & $<0.001$ & 0.022 \\
    344   & $<0.001$ & 0.047 & 0.033 & $<0.001$ & 0.080 & $<0.001$ & 0.040 & 0.140 & $<0.001$ & $<0.001$ \\
    411.8 & $<0.001$ & $<0.001$ & $<0.001$ & $<0.001$ & 0.385 & $<0.001$ & $<0.001$ & $<0.001$ & $<0.001$ & $<0.001$ \\
    443.96 & $<0.001$ & $<0.001$ & $<0.001$ & $<0.001$ & 0.053 & $<0.001$ & $<0.001$ & 0.111 & 0.012 & $<0.001$ \\
    465   & 0.758 & 0.542 & 0.518 & 0.488 & 0.285 & 0.120 & 0.535 & 0.302 & 0.287 & 0.290 \\
    468.1 & $<0.001$ & $<0.001$ & $<0.001$ & $<0.001$ & 0.051 & $<0.001$ & $<0.001$ & $<0.001$ & $<0.001$ & $<0.001$ \\
    511   & 0.754 & 0.752 & 0.756 & 0.756 & 0.730 & 0.746 & 0.753 & 0.727 & 0.731 & 0.731 \\
    661.6 & $<0.001$ & $<0.001$ & $<0.001$ & $<0.001$ & 0.028 & $<0.001$ & $<0.001$ & $<0.001$ & $<0.001$ & $<0.001$ \\
    723   & $<0.001$ & $<0.001$ & $<0.001$ & $<0.001$ & 0.096 & $<0.001$ & $<0.001$ & 0.114 & 0.009 & $<0.001$ \\
    779   & $<0.001$ & $<0.001$ & $<0.001$ & $<0.001$ & 0.683 & $<0.001$ & $<0.001$ & 0.333 & $<0.001$ & $<0.001$ \\
    811   & 0.102 & 0.048 & 0.044 & 0.043 & 0.054 & 0.012 & 0.046 & 0.054 & 0.053 & 0.055 \\
    847   & $<0.001$ & $<0.001$ & $<0.001$ & $<0.001$ & 0.598 & $<0.001$ & $<0.001$ & 0.870 & 0.278 & $<0.001$ \\
    868   & 0.005 & 0.162 & 0.141 & 0.043 & 0.919 & $<0.001$ & 0.142 & 0.925 & 0.854 & $<0.001$ \\
    878   & 0.026 & 0.140 & 0.119 & 0.115 & 0.164 & 0.001 & 0.123 & 0.153 & 0.140 & 0.156 \\
    889   & $<0.001$ & $<0.001$ & $<0.001$ & $<0.001$ & 0.042 & $<0.001$ & $<0.001$ & $<0.001$ & $<0.001$ & $<0.001$ \\
    952   & $<0.001$ & $<0.001$ & $<0.001$ & $<0.001$ & $<0.001$ & $<0.001$ & $<0.001$ & $<0.001$ & $<0.001$ & $<0.001$ \\
    964   & $<0.001$ & $<0.001$ & $<0.001$ & $<0.001$ & 0.152 & $<0.001$ & $<0.001$ & $<0.001$ & $<0.001$ & $<0.001$ \\
    1005  & $<0.001$ & $<0.001$ & $<0.001$ & $<0.001$ & 0.987 & $<0.001$ & $<0.001$ & 0.968 & 0.697 & $<0.001$ \\
    1019  & 0.280 & 0.053 & 0.045 & 0.045 & 0.075 & $<0.001$ & 0.049 & 0.073 & 0.066 & 0.095 \\
    1038  & $<0.001$ & 0.041 & 0.011 & 0.011 & 0.004 & $<0.001$ & 0.056 & 0.019 & $<0.001$ & $<0.001$ \\
    1085.87 & $<0.001$ & $<0.001$ & $<0.001$ & $<0.001$ & 0.125 & $<0.001$ & $<0.001$ & 0.011 & $<0.001$ & $<0.001$ \\
    1112  & $<0.001$ & $<0.001$ & $<0.001$ & $<0.001$ & 0.060 & $<0.001$ & $<0.001$ & $<0.001$ & $<0.001$ & $<0.001$ \\
    1120  & $<0.001$ & $<0.001$ & $<0.001$ & $<0.001$ & 0.090 & $<0.001$ & $<0.001$ & $<0.001$ & $<0.001$ & $<0.001$ \\
    1173  & $<0.001$ & $<0.001$ & $<0.001$ & $<0.001$ & 0.090 & $<0.001$ & $<0.001$ & $<0.001$ & $<0.001$ & $<0.001$ \\
    1189  & 0.948 & 0.401 & 0.380 & 0.380 & 0.523 & 0.022 & 0.384 & 0.482 & 0.468 & 0.522 \\
    1238  & $<0.001$ & $<0.001$ & $<0.001$ & $<0.001$ & 0.011 & $<0.001$ & $<0.001$ & 0.008 & $<0.001$ & $<0.001$ \\
    1274  & $<0.001$ & $<0.001$ & $<0.001$ & $<0.001$ & 0.076 & $<0.001$ & $<0.001$ & 0.016 & $<0.001$ & $<0.001$ \\
    1302  & 0.054 & 0.039 & 0.031 & 0.031 & 0.050 & $<0.001$ & 0.032 & 0.048 & 0.044 & 0.057 \\
    1332  & $<0.001$ & $<0.001$ & $<0.001$ & $<0.001$ & $<0.001$ & $<0.001$ & $<0.001$ & $<0.001$ & $<0.001$ & $<0.001$ \\
    1359  & $<0.001$ & $<0.001$ & $<0.001$ & $<0.001$ & 0.382 & $<0.001$ & $<0.001$ & 0.325 & 0.133 & $<0.001$ \\
    1408  & $<0.001$ & $<0.001$ & $<0.001$ & $<0.001$ & 0.312 & $<0.001$ & $<0.001$ & 0.068 & $<0.001$ & $<0.001$ \\
    $\geq$1600 & $<0.001$ & $<0.001$ & $<0.001$ & $<0.001$ & $<0.001$ & $<0.001$ & $<0.001$ & $<0.001$ & $<0.001$ & $<0.001$ \\
    \end{tabular}%
  \label{tab_pvalue_90}%
\end{table*}%

\begin{table*}[htbp]
  \centering
  \caption{Differential cross sections: p-values resulting from the $\chi^{2}$ test, data sample above 90$^{\circ}$}
    \begin{tabular}{c|rrrrrrrrrr}
    \hline
    \textbf{E } & {\textbf{Penelope}} & {\textbf{Penelope}} & {\textbf{EPDL}} & {\textbf{EPDL}} & {\textbf{SM}} & {\textbf{RF}} & {\textbf{NF}} & {\textbf{MF}} & {\textbf{MF}} & {\textbf{RF}} \\
    (keV) & {\textbf{2001}} & {\textbf{2008-2011}} & {\textbf{}} & {\textbf{ASF}} & {} & {\textbf{}} & {\textbf{}} & {\textbf{}} & {\textbf{ASF}} & {\textbf{ASF}} \\
\hline
    13.95 & $<0.001$ & $<0.001$ & $<0.001$ & 0.002 & 0.154 & $<0.001$ & $<0.001$ & $<0.001$ & 0.046 & 0.064 \\
    17.75 & $<0.001$ & $<0.001$ & $<0.001$ & $<0.001$ & 0.105 & $<0.001$ & $<0.001$ & $<0.001$ & 0.009 & 0.025 \\
    22.1  & $<0.001$ & $<0.001$ & $<0.001$ & $<0.001$ & 0.010 & $<0.001$ & $<0.001$ & $<0.001$ & $<0.001$ & $<0.001$ \\
    26.36 & $<0.001$ & $<0.001$ & $<0.001$ & $<0.001$ & $<0.001$ & $<0.001$ & $<0.001$ & $<0.001$ & 0.772 & 0.758 \\
    59.54 & $<0.001$ & $<0.001$ & $<0.001$ & $<0.001$ & $<0.001$ & $<0.001$ & $<0.001$ & $<0.001$ & $<0.001$ & 0.001 \\
    81    & $<0.001$ & $<0.001$ & $<0.001$ & $<0.001$ & 0.365 & $<0.001$ & $<0.001$ & $<0.001$ & 0.004 & 0.082 \\
    88    & $<0.001$ & $<0.001$ & $<0.001$ & $<0.001$ & $<0.001$ & $<0.001$ & $<0.001$ & $<0.001$ & $<0.001$ & $<0.001$ \\
    145.4 & $<0.001$ & $<0.001$ & $<0.001$ & $<0.001$ & $<0.001$ & $<0.001$ & $<0.001$ & $<0.001$ & $<0.001$ & $<0.001$ \\
    279.2 & $<0.001$ & $<0.001$ & $<0.001$ & $<0.001$ & 0.096 & $<0.001$ & $<0.001$ & $<0.001$ & $<0.001$ & $<0.001$ \\
    317   & $<0.001$ & $<0.001$ & $<0.001$ & 0.019 & $<0.001$ & 0.200 & $<0.001$ & $<0.001$ & $<0.001$ & $<0.001$ \\
    411.8 & $<0.001$ & $<0.001$ & $<0.001$ & $<0.001$ & 0.034 & $<0.001$ & $<0.001$ & $<0.001$ & $<0.001$ & $<0.001$ \\
    468.1 & 0.015 & 0.047 & 0.037 & $<0.001$ & 0.368 & $<0.001$ & 0.045 & $<0.001$ & $<0.001$ & $<0.001$ \\
    661.6 & $<0.001$ & $<0.001$ & $<0.001$ & $<0.001$ & 0.508 & $<0.001$ & $<0.001$ & $<0.001$ & $<0.001$ & $<0.001$ \\
    889   & $<0.001$ & $<0.001$ & $<0.001$ & $<0.001$ & 0.940 & $<0.001$ & $<0.001$ & $<0.001$ & $<0.001$ & $<0.001$ \\
    1120  & $<0.001$ & $<0.001$ & $<0.001$ & $<0.001$ & $<0.001$ & $<0.001$ & $<0.001$ & $<0.001$ & $<0.001$ & $<0.001$ \\
    1173  & $<0.001$ & $<0.001$ & $<0.001$ & $<0.001$ & 0.022 & $<0.001$ & $<0.001$ & $<0.001$ & $<0.001$ & $<0.001$ \\
    1332  & $<0.001$ & $<0.001$ & $<0.001$ & $<0.001$ & $<0.001$ & $<0.001$ & $<0.001$ & $<0.001$ & $<0.001$ & $<0.001$ \\
\hline
    \end{tabular}%
  \label{tab_pvalue_gt90}%
\end{table*}%

The efficiency of the various simulation models is reported in Table
\ref{tab_effall}.
The results are listed for three data samples: the whole experimental data
collection, the data for scattering angles up to $90^{\circ}$ and above
$90^{\circ}$.
The simulation model based on the interpolation of RTAB S-matrix calculations,
indicated in the tables as SM,
stands out as the one exhibiting the largest efficiency in all test
configurations.

Three test cases for which the $\chi^{2}$ test rejects the hypothesis of
compatibility between the S-matrix model and experimental data involve
measurements from a single reference; due to the lack of other experimental
data in the same configuration, one cannot exclude that the
incompatibility between S-matrix calculations and experiment could be ascribed
to systematic effects in the data, rather than to deficiencies of the simulation
model.
The exclusion of these three test cases from the analysis slightly increases the
efficiencies of all the simulation models (e.g. 0.81$\pm$0.05 for the S-matrix
model and 0.40$\pm$0.05 for the EPDL-based model, over the whole angular range),
nevertheless it does not substantially change the conclusions one can draw from
the $\chi^{2}$ test.

Several cases of inconsistency between simulation models and experimental data
concern measurements at photon energies close to the K or L shell 
electron binding energies of target atom; a number of them involve 
measurements for which relatively small experimental uncertainties  
are reported, that could have been underestimated.
The same analysis has also been performed on a sample that excludes these 
data; its results, which are reported in Table \ref{tab_effbinding}, confirm the
same general trend observed in Table \ref{tab_effall}.
The larger relative increase in efficiency of the models based on RTAB observed
in this analysis with respect to the effect on other models suggests that this
database might have adopted a coarse grid for its tabulations, that degrades the
accuracy of interpolation in sensitive areas.

The compatibility between simulation models and experimental data has been
evaluated also through the Kolmogorov-Smirnov, Anderson-Darling and Cramer-von
Mises goodness-of-fit tests.
The results, listed in Table \ref{tab_gof}, suggest that these tests exhibit scarce
sensitivity to differences between the simulation models and the experimental data
subject to comparison.

\begin{table*}[htbp]
  \centering
  \caption{$\chi^{2}$ test outcome: test cases compatible with experiment at 0.01 significance level}
    \begin{tabular}{c|l|cccccccccc}
    \hline
    {\textbf{Test}} &       & {\textbf{Penelope}} & {\textbf{Penelope}} & {\textbf{EPDL}} & {\textbf{EPDL}} & {\textbf{SM}} & {\textbf{RF}} & {\textbf{NF}} & {\textbf{MF}} & {\textbf{MF}} & {\textbf{RF}} \\
          &       & {\textbf{2001}} & {\textbf{2008-2011}} & {\textbf{}} & {\textbf{ASF}} & {} & {\textbf{}} & {\textbf{}} & {\textbf{}} & {\textbf{ASF}} & {\textbf{ASF}} \\
    \hline
    \multirow{5}[2]{*}{all} & Test cases & 71    & 71    & 71    & 71    & 71    & 71    & 71    & 71    & 71    & 71 \\
    & Pass  & 19    & 27    & 27    & 18    & 55    & 18    & 25    & 35    & 37    & 34 \\
    & Fail    & 52    & 44    & 44    & 53    & 16    & 53    & 46    & 36    & 34    & 37 \\
    & Efficiency & 0.27  & 0.38  & 0.38  & 0.25  & 0.77  & 0.25  & 0.35  & 0.49  & 0.52  & 0.48 \\
          & Error & $\pm$0.05  & $\pm$0.06  & $\pm$0.06  & $\pm$0.05  & $\pm$0.06  & $\pm$0.05  & $\pm$0.06  & $\pm$0.06  & $\pm$0.06  & $\pm$0.06 \\
    \hline
    \multirow{5}[2]{*}{$\theta\leq90^{\circ}$} & Test cases & 67    & 67    & 67    & 67    & 67    & 67    & 67    & 67    & 67    & 67 \\
          & Pass  & 19    & 27    & 27    & 18    & 55    & 18    & 25    & 35    & 36    & 32 \\
          & Fail    & 48    & 40    & 40    & 49    & 12    & 49    & 42    & 32    & 31    & 35 \\
          & Efficiency & 0.28  & 0.40  & 0.40  & 0.27  & 0.82  & 0.27  & 0.37  & 0.52  & 0.54  & 0.48 \\
          & Error & $\pm$0.05  & $\pm$0.06  & $\pm$0.06  & $\pm$0.05  & $\pm$0.05  & $\pm$0.05  & $\pm$0.06  & $\pm$0.06  & $\pm$0.06  & $\pm$0.06 \\
\hline  
  \multirow{5}[2]{*}{$\theta>90^{\circ}$}  & Test cases & 17    & 17    & 17    & 17    & 17    & 17    & 17    & 17    & 17    & 17 \\
          & Pass  & 1     & 1     & 1     & 1     & 10     & 1     & 1     & 0     & 2     & 4 \\
          & Fail  & 16    & 16    & 16    & 16    & 7    & 16    & 16    & 17    & 15    & 13 \\
          & Efficiency & 0.06  & 0.06  & 0.06  & 0.06  & 0.59  & 0.06  & 0.06  & $<0.06$  & 0.12  & 0.24 \\
          & Error & $\pm$0.06  & $\pm$0.06  & $\pm$0.06  &$\pm$0.06   & $\pm$0.12  & $\pm$0.06  & $\pm$0.06  &   & $\pm$0.08  & $\pm$0.10 \\
    \hline
    \end{tabular}%
  \label{tab_effall}%
\end{table*}%

\begin{table*}[htbp]
  \centering
  \caption{$\chi^{2}$ test outcome, excluding data sensitive to K and L shell effects: test cases compatible with experiment at 0.01 significance level}
    \begin{tabular}{c|l|cccccccccc}
    \hline
    {\textbf{Test}} &       & {\textbf{Penelope}} & {\textbf{Penelope}} & {\textbf{EPDL}} & {\textbf{EPDL}} & {\textbf{SM}} & {\textbf{RF}} & {\textbf{NF}} & {\textbf{MF}} & {\textbf{MF}} & {\textbf{RF}} \\
          &       & {\textbf{2001}} & {\textbf{2008-2011}} & {\textbf{}} & {\textbf{ASF}} & {} & {\textbf{}} & {\textbf{}} & {\textbf{}} & {\textbf{ASF}} & {\textbf{ASF}} \\

\hline
    \multirow{5}[2]{*}{all} & Test cases & 71    & 71    & 71    & 71    & 71    & 71    & 71    & 71    & 71    & 71 \\
    & Pass  	& 19    & 28    & 28    & 18   & 62    & 21    & 26    & 36    & 43    & 40 \\
    &Fail  	& 52    & 43    & 43    & 53    & 9    & 50    & 45    & 35    & 28    & 31 \\
    &Efficiency & 0.27  & 0.39  & 0.39  & 0.35  & 0.87  & 0.30  & 0.37  & 0.51  & 0.61  & 0.56 \\
    & Error 	& $\pm$0.05  & $\pm$0.06  & $\pm$0.06  & $\pm$0.06  & $\pm$0.04  & $\pm$0.05  & $\pm$0.06  & $\pm$0.06  & $\pm$0.06  & $\pm$0.06 \\
    \hline
    \multirow{5}[2]{*}{$\theta\leq90^{\circ}$} & Test cases & 67    & 67    & 67    & 67    & 67    & 67    & 67    & 67    & 67    & 67 \\
     & Pass  		& 19    & 28    & 28    & 25    & 61    & 21    & 26    & 36    & 42    & 38 \\
     & Fail  		& 48    & 39    & 39    & 42    & 6     & 46    & 41    & 31    & 25    & 29 \\
    & Efficiency 	& 0.28  & 0.42  & 0.42  & 0.37  & 0.91  & 0.31  & 0.39  & 0.54  & 0.63  & 0.57 \\
       & Error 		& $\pm$0.06  & $\pm$0.06  & $\pm$0.06  & $\pm$0.06  & $\pm$0.04  & $\pm$0.06  & $\pm$0.06  & $\pm$0.06  & $\pm$0.06  & $\pm$0.06 \\
    \hline
    \multirow{5}[2]{*}{$\theta>90^{\circ}$} & Test cases & 17    & 17    & 17    & 17    & 17    & 17    & 17    & 17    & 17    & 17 \\
          & Pass  	 & 1     & 1     & 1     & 1     & 11    & 1     & 1     & 0     & 3     & 4 \\
          & Fail  	 & 16    & 16    & 16    & 16    & 6     & 16    & 16    & 17    & 14    & 13 \\
          & Efficiency & 0.06  & 0.06  & 0.06  & 0.06  & 0.65  & 0.06  & 0.06  & $<0.06$  & 0.18  & 0.24 \\
           & Error 	& $\pm$0.06  & $\pm$0.06  & $\pm$0.06  &$\pm$0.06   & $\pm$0.12  & $\pm$0.06  & $\pm$0.06  &    & $\pm$0.09  & $\pm$0.10 \\
    \hline
    \end{tabular}%
  \label{tab_effbinding}%
\end{table*}%


\begin{table*}[htbp]
  \centering
  \caption{Goodness-of-fit tests: test cases compatible with experiment at 0.05 significance level}
    \begin{tabular}{c|l|cccccccccc}
    \hline
     {\textbf{Test}} &       & {\textbf{Penelope}} & {\textbf{Penelope}} & {\textbf{EPDL}} & {\textbf{EPDL}} & {\textbf{SM}} & {\textbf{RF}} & {\textbf{NF}} & {\textbf{MF}} & {\textbf{MF}} & {\textbf{RF}} \\
          &       & {\textbf{2001}} & {\textbf{2008-2011}} & {\textbf{}} & {\textbf{ASF}} & {} & {\textbf{}} & {\textbf{}} & {\textbf{}} & {\textbf{ASF}} & {\textbf{ASF}} \\
\hline
          & {Test cases} & 71    & 71    & 71    & 71    & 71    & 71    & 71    & 71    & 71    & 71 \\
    \hline
   & Pass  & 67    & 68    & 68    & 61    & 69    & 62    & 67    & 62    & 69    & 64 \\
 Kolmogorov   & Fail  & 4     & 3     & 3     & 10    & 2     & 9     & 4     & 9     & 2     & 7 \\  
     Smirnov      & Efficiency & 0.94  & 0.96  & 0.96  & 0.86  & 0.97  & 0.87  & 0.94  & 0.87  & 0.97  & 0.90 \\
          & Error & $\pm$0.03  & $\pm$0.02  & $\pm$0.02  & $\pm$0.04  & $\pm$0.02  & $\pm$0.04  & $\pm$0.03  & $\pm$0.04  & $\pm$0.02  & $\pm$0.04\\
    \hline
     & Pass  & 68    & 69    & 69    & 62    & 70    & 63    & 69    & 65    & 70    & 63 \\
Anderson    & Fail  & 3     & 2     & 2     & 9     & 1     & 8     & 2     & 6     & 1     & 8 \\
     Darling      & Efficiency & 0.96  & 0.97  & 0.97  & 0.87  & 0.99  & 0.89  & 0.97  & 0.92  & 0.99  & 0.89 \\
          & Error & $\pm$0.02  & $\pm$0.02  & $\pm$0.02  & $\pm$0.04  & $\pm$0.01  & $\pm$0.04  & $\pm$0.02  & $\pm$0.03  & $\pm$0.01  & $\pm$0.04 \\
    \hline
   & Pass  & 69    & 69    & 69    & 62    & 70    & 66    & 69    & 67    & 70    & 63 \\
  Cramer    &Fail  & 2     & 2     & 2     & 9     & 1     & 5     & 2     & 4     & 1     & 8 \\
        von Mises   & Efficiency & 0.97  & 0.97  & 0.97  & 0.87  & 0.99  & 0.93  & 0.97  & 0.94  & 0.99  & 0.89 \\
          & Error & $\pm$0.02  & $\pm$0.02  & $\pm$0.02  & $\pm$0.04  & $\pm$0.01  & $\pm$0.03  & $\pm$0.02  & $\pm$0.03  & $\pm$0.01  & $\pm$0.04 \\
    \hline
    \end{tabular}%
  \label{tab_gof}%
\end{table*}%

The relative accuracy of the various simulation models  is compared by
contingency tables based on the outcome of the $\chi^2$ test.
In these comparisons the model based on the S-matrix, which exhibits the
largest efficiency, is taken as a reference; contingency tables examine whether
the results of the  $\chi^2$ test over other models are statistically equivalent to
those of this model.

Since several general purpose Monte Carlo systems base their simulation of
photon elastic scattering on EPDL, it is interesting to quantify whether the
differential cross sections based on this data library differ significantly in compatibility
with experiment from those derived from S-matrix calculations.
This analysis is summarized in Table \ref{tab_contall}, which reports results
for the whole data sample, and in Table \ref{tab_contbind}, which excludes data
at energies close to K and L shell binding energies;
both tables list results calculated over the whole data sample and for
scattering angles up to, and above 90$^{\circ}$.
The hypothesis of statistically equivalent capability to reproduce experimental
measurements by the S-matrix simulation model and by the model based on EPDL is
rejected at 0.01 level of significance.

Tables \ref{tab_contall} and \ref{tab_contbind} also report a comparison between
the S-matrix model and the model exploiting modified form factors with anomalous
scattering factors (MFASF), which exhibits the highest efficiencies in Tables
\ref{tab_effall} and \ref{tab_effbinding} among those based on the form factor
approximation.
The hypothesis of equivalent accuracy is rejected with 0.01 significance also
for the MFASF model over the whole data sample and the subset of data with
scattering angles up to 90$^{\circ}$, while it is not rejected for the
backscattering data sample, although with p-values close to the critical region.
A fortiori, simulation models exhibiting lower efficiencies than the MFASF model
in Table \ref{tab_contall} are significantly less accurate than the model based
on RTAB S-matrix calculations.

From this analysis one can draw the conclusion that the simulation model based
on RTAB tabulations of S-matrix calculations ensures significantly more accurate
results than all the other simulation alternatives considered in this study,
including the EPDL data library currently used by several general purpose Monte
Carlo systems.
The model exploiting modified form factors with anomalous scattering factors is
the most accurate among those based on the form factor approximation.

The model based on pristine EPDL97 tabulations of non-relativistic form factors
and the Penelope 2008-2011 model, based on modified tabulations derived from
EPDL97, produce equivalent differential cross sections: their efficiencies
listed in Table \ref{tab_contall} and \ref{tab_contbind} are identical, despite
the small differences visible in Fig. \ref{fig_penelope}.
The tabulations of non-relativistic form factors included in RTAB produce
slightly less accurate cross sections than the models based on EPDL97, although
the differences in efficiencies are compatible within one standard deviation.

The simulation model based on relativistic form factors is less accurate than models 
exploiting non-relativistic form factors: 
it has been previously noted \cite{Kane1983, Roy1983} that the use of
relativistic wavefunctions in the calculation of form factors often produces
less accurate results than use of nonrelativistic wavefunctions, although - to
the best of our knowledge - the relative efficiency of these two calcluation
methods at reproducing experimental data has not been yet quantified with
statistical methods.

The Penelope 2001 model is less accurate than more recent versions of the code.

The inclusion of anomalous scattering factors in the calculations based on
EPDL97 does not contribute to improve compatibility with experiment, while
accounting for anomalous scattering improves the compatibility with experiment
of calculations exploiting relativistic and modified form factors.

\begin{table}[htbp]
  \centering
  \caption{Contingency tables related to the whole data sample}
    \begin{tabular}{|c|rrr|}
    \hline
    \multirow{14}[4]{*}{All angles} &       & \textbf{SM} & \textbf{EPDL} \\
          & Pass  & 55    & 27 \\
          & Fail  & 16    & 44 \\
          &       & \multicolumn{2}{c|}{\textbf{p-value}}  \\
          & Fisher test & \multicolumn{2}{c|}{$<0.001$}  \\
          & Pearson's $\chi^2$ & \multicolumn{2}{c|}{$<0.001$}  \\
          & Yates $\chi^2$ & \multicolumn{2}{c|}{$<0.001$}  \\
\cline{2-4}          &       & \textbf{SM} & \textbf{MFASF} \\
          & Pass  & 55    & 37 \\
          & Fail  & 16    & 34 \\
          &       & \multicolumn{2}{c|}{\textbf{p-value}}  \\
          & Fisher test & \multicolumn{2}{c|}{0.003}  \\
          & Pearson's $\chi^2$ & \multicolumn{2}{c|}{0.002} \\
          & Yates $\chi^2$ & \multicolumn{2}{c|}{0.002}  \\
    \hline
    \multirow{14}[4]{*}{$\theta\leq90^{\circ}$} &       & \textbf{SM} & \textbf{EPDL} \\
          & Pass  & 55    & 27 \\
          & Fail  & 12    & 40 \\
          &       & \multicolumn{2}{c|}{\textbf{p-value}}  \\
          & Fisher test & \multicolumn{2}{c|}{$<0.001$}  \\
          & Pearson's $\chi^2$ & \multicolumn{2}{c|}{$<0.001$}  \\
          & Yates $\chi^2$ & \multicolumn{2}{c|}{$<0.001$}  \\
\cline{2-4}          &       & \textbf{SM} & \textbf{MFASF} \\
          & Pass  & 55    & 36 \\
          & Fail  & 12    & 31 \\
          &       & \multicolumn{2}{c|}{\textbf{p-value}}  \\
          & Fisher test & \multicolumn{2}{c|}{$<0.001$} \\
          & Pearson's $\chi^2$ & \multicolumn{2}{c|}{$<0.001$}  \\
          & Yates $\chi^2$ & \multicolumn{2}{c|}{$<0.001$}  \\
    \hline
    \multirow{14}[4]{*}{$\theta > 90^{\circ}$} &       & \textbf{SM} & \textbf{EPDL} \\
          & Pass  & 10    & 1 \\
          & Fail  & 7    & 16 \\
          &       & \multicolumn{2}{c|}{\textbf{p-value}}  \\
          & Fisher test & \multicolumn{2}{c|}{0.002}  \\
          & Yates $\chi^2$ & \multicolumn{2}{c|}{0.001}  \\
\cline{2-4}          &       & \textbf{SM} & \textbf{MFASF} \\
          & Pass  & 10    & 2 \\
          & Fail  & 7    & 15 \\
          &       & \multicolumn{2}{c|}{\textbf{p-value}}  \\
          & Fisher test & \multicolumn{2}{c|}{0.010} \\
          & Yates $\chi^2$ & \multicolumn{2}{c|}{0.012}  \\
    \hline
    \end{tabular}%
  \label{tab_contall}%
\end{table}%

\begin{table}[htbp]
  \centering
  \caption{Contingency tables excluding energies close to K and L shell binding energies}
    \begin{tabular}{|c|rrr|}
    \hline
    \multirow{14}[4]{*}{All angles} &       & \textbf{SM} & \textbf{EPDL} \\
          & Pass  & 62    & 28 \\
          & Fail  & 9    & 43 \\
          &       & \multicolumn{2}{c|}{\textbf{p-value}} \\
          & Fisher test & \multicolumn{2}{c|}{ $<0.001$} \\
          & Pearson's $\chi^2$ & \multicolumn{2}{c|}{ $<0.001$} \\
          & Yates $\chi^2$ & \multicolumn{2}{c|}{ $<0.001$}  \\
\cline{2-4}          &       & \textbf{SM} & \textbf{MFASF} \\
          & Pass  & 62    & 43 \\
          & Fail  & 9    & 28 \\
          &       & \multicolumn{2}{c|}{\textbf{p-value}}  \\
          & Fisher test & \multicolumn{2}{c|}{$<0.001$}  \\
          & Pearson's $\chi^2$ & \multicolumn{2}{c|}{$<0.001$}  \\
          & Yates $\chi^2$ & \multicolumn{2}{c|}{$<0.001$}  \\
    \hline
    \multirow{14}[4]{*}{$\theta\leq90^{\circ}$} &       & \textbf{SM} & \textbf{EPDL} \\
          & Pass  & 61    & 28 \\
          & Fail  & 6     & 39 \\
          &       & \multicolumn{2}{c|}{\textbf{p-value}}  \\
          & Fisher test & \multicolumn{2}{c|}{ $<0.001$}  \\
          & Pearson's $\chi^2$ & \multicolumn{2}{c|}{ $<0.001$}  \\
          & Yates $\chi^2$ & \multicolumn{2}{c|}{ $<0.001$}  \\
\cline{2-4}          &       & \textbf{SM} & \textbf{MFASF} \\
          & Pass  & 61    & 42 \\
          & Fail  & 6     & 25 \\
          &       & \multicolumn{2}{c|}{\textbf{p-value}}  \\
          & Fisher test & \multicolumn{2}{c|}{$<0.001$}  \\
          & Pearson's $\chi^2$ & \multicolumn{2}{c|}{$<0.001$}  \\
          & Yates $\chi^2$ & \multicolumn{2}{c|}{$<0.001$}  \\
    \hline
    \multirow{14}[4]{*}{$\theta > 90^{\circ}$} &       & \textbf{SM} & \textbf{EPDL} \\
          & Pass  & 11    & 1 \\
          & Fail  & 6     & 16 \\
          &       & \multicolumn{2}{c|}{\textbf{p-value}}  \\
          & Fisher test & \multicolumn{2}{c|}{ $<0.001$}  \\
          & Yates $\chi^2$ & \multicolumn{2}{c|}{ 0.001}  \\
\cline{2-4}          &       & \textbf{SM} & \textbf{MFASF} \\
          & Pass  & 11    & 3 \\
          & Fail  & 6     & 4 \\
          &       & \multicolumn{2}{c|}{\textbf{p-value}}  \\
          & Fisher test & \multicolumn{2}{c|}{0.013}  \\
          & Yates $\chi^2$ & \multicolumn{2}{c|}{0.015}  \\
    \hline

    \end{tabular}%
  \label{tab_contbind}%
\end{table}%

\begin{figure}
\centerline{\includegraphics[angle=0,width=8.5cm]{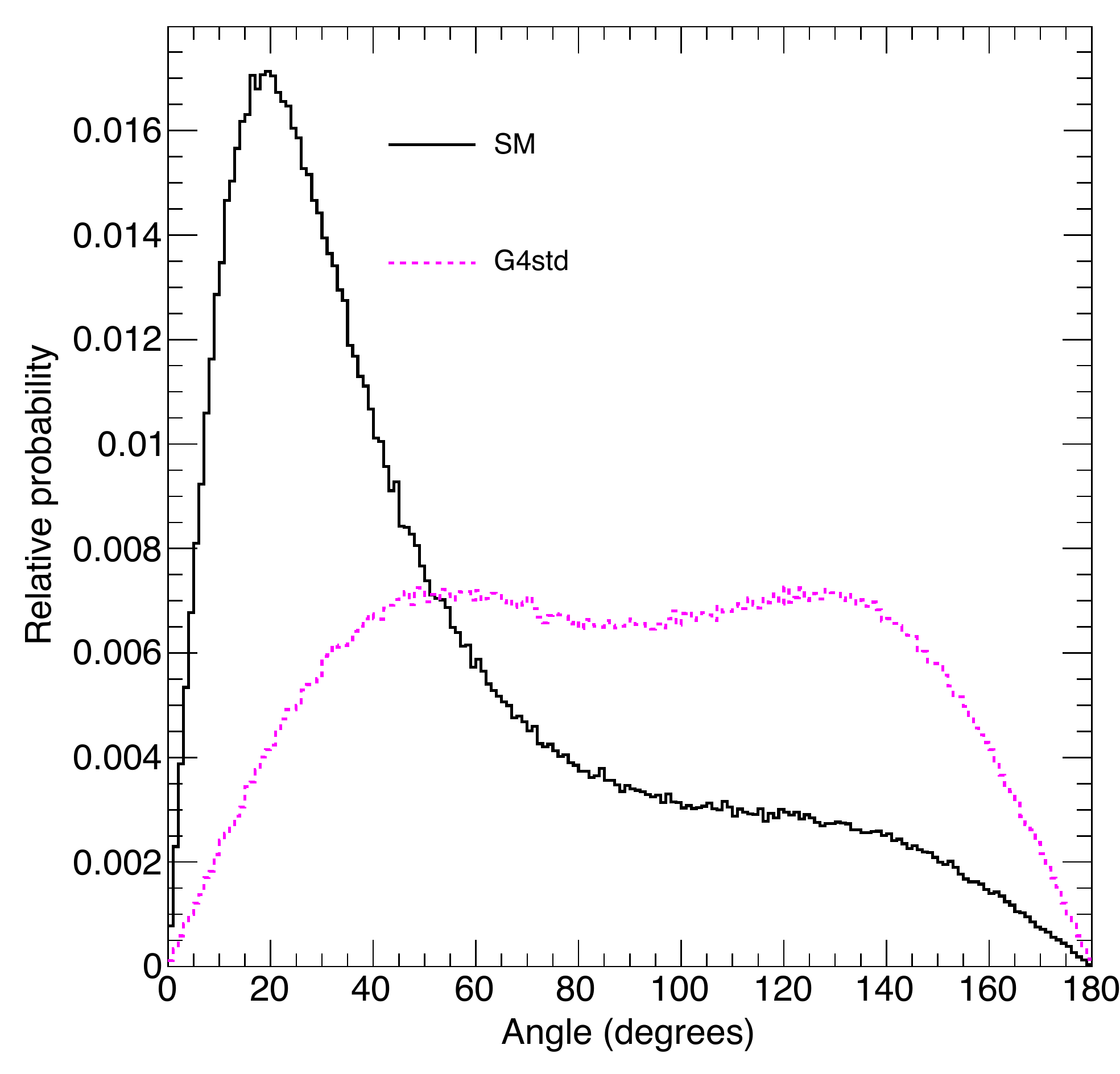}}
\caption{Probability distribution of the scattering angle of 10~keV photons
interacting with carbon generated by \textit{G4XrayRayleighModel} (magenta
dashed line) this model, compared to the distribution based on S-matrix
calculations (black solid line; both distributions are normalized to 1. 
The distribution of \textit{G4XrayRayleighModel} corresponds to scattering from
a point-like charge.}
\label{fig_stdangle}
\end{figure}

The simplified model in \textit{G4XrayRayleighModel}, which implements elastic
scattering from a point-like charge, produces a largely unrealistic description
of photon elastic scattering: this is visible, for instance, in Fig.
\ref{fig_stdangle}, which shows the distribution of the scattering angle of
10~keV photons interacting with carbon generated by this model, compared to the
distribution based on S-matrix calculations.
Given its evident inadequacy, this model was not included in the statistical 
analysis for the validation of differential cross sections.

\subsection{Total Cross Sections}
\label{sec_totcs}

\begin{figure*}
\centerline{\includegraphics[angle=0,width=18cm]{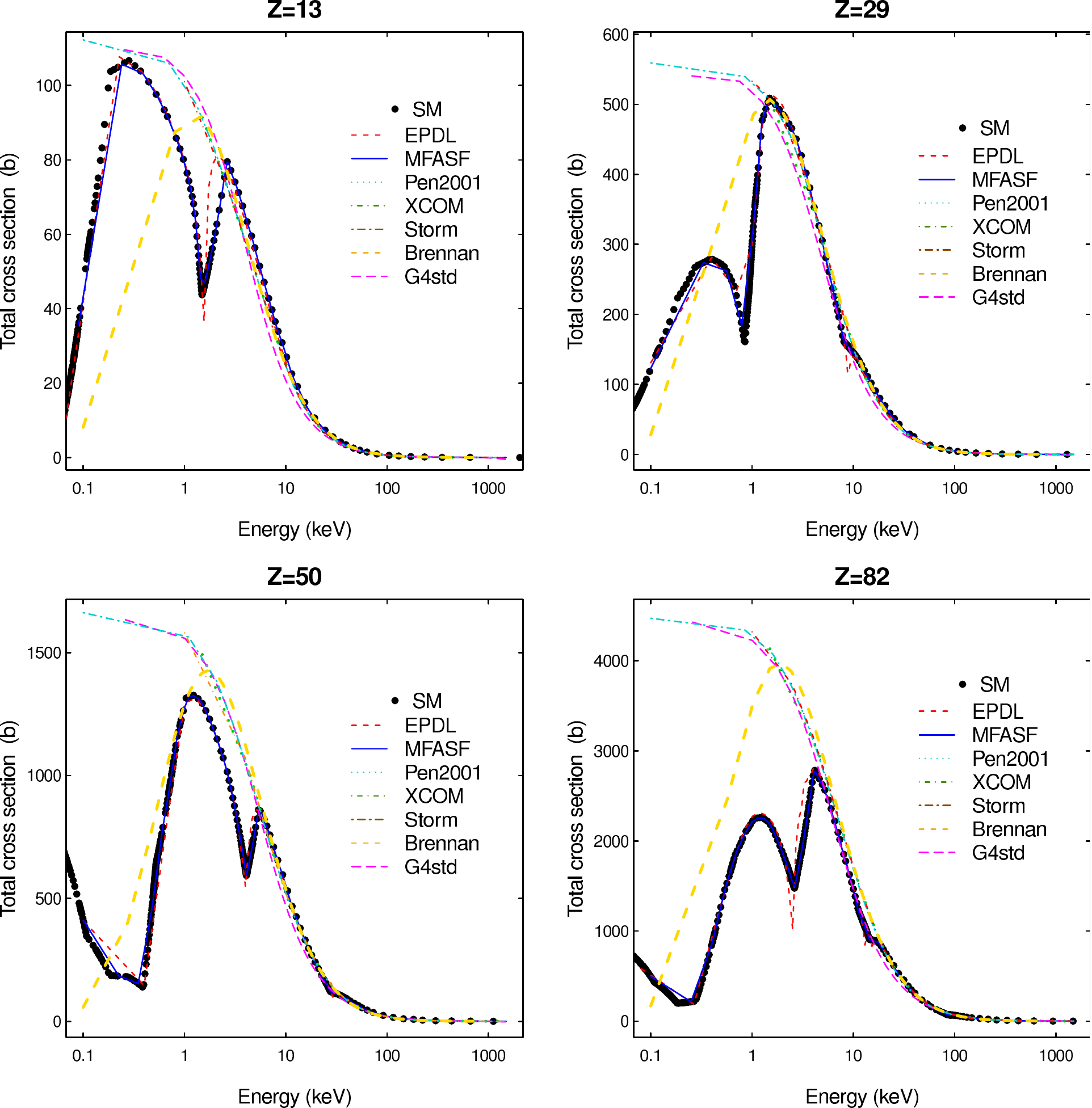}}
\caption{Total cross section as a function of energy for aluminium, copper, tin
and lead: calculations based on S-matrix (SM, black filled circles), EPDL (red
dashed line), modified form factors with anomalous scattering factors MFASF
(blue solid line), XCOM (green dotted line), Storm and Israel (brown
dotted-dashed line), Penelope 2001 (turquoise double-dashed line),
Brennan and Cowan (yellow dashed line)  and 
\textit{G4XrayRayleighModel} (G4std, magenta long-dashed line). }
\label{fig_totcs}
\end{figure*}

Total cross sections calculated by a set of simulation models are shown in Fig.
\ref{fig_totcs} for four representative elements.
Some of them are associated with modeling approaches whose capabilities have
been evaluated in the previous section regarding differential cross sections:
the models based on S-matrix calculations, EPDL97, modified form factors with
anomalous scattering factors and Penelope 2001 parameterizations.
Others (XCOM, Storm and Israel, Brennan and Cowan), are specific for total cross
section calculation only.
In addition, the most recent development in the field, Geant4
\textit{G4XrayRayleighModel}, is illustrated in the plots.

The total cross sections calculated by XCOM, Penelope 2001, Storm and
Israel, and \textit{G4XrayRayleighModel} appear
insensitive to the underlying atomic structure; for photon energies below 10 keV
these models overestimate the total cross section with respect to the
S-matrix model, which is the result of the integration of the differential cross
sections assessed in section \ref{sec_diffcs} as best reproducing
experimental data.
The parameterized cross sections by Brennan and Cowan attempt to account for
atomic shell effects at low energies, but their results appear largely
approximate with respect to S-matrix calculations.

\begin{figure*}
\centerline{\includegraphics[angle=0,width=18cm]{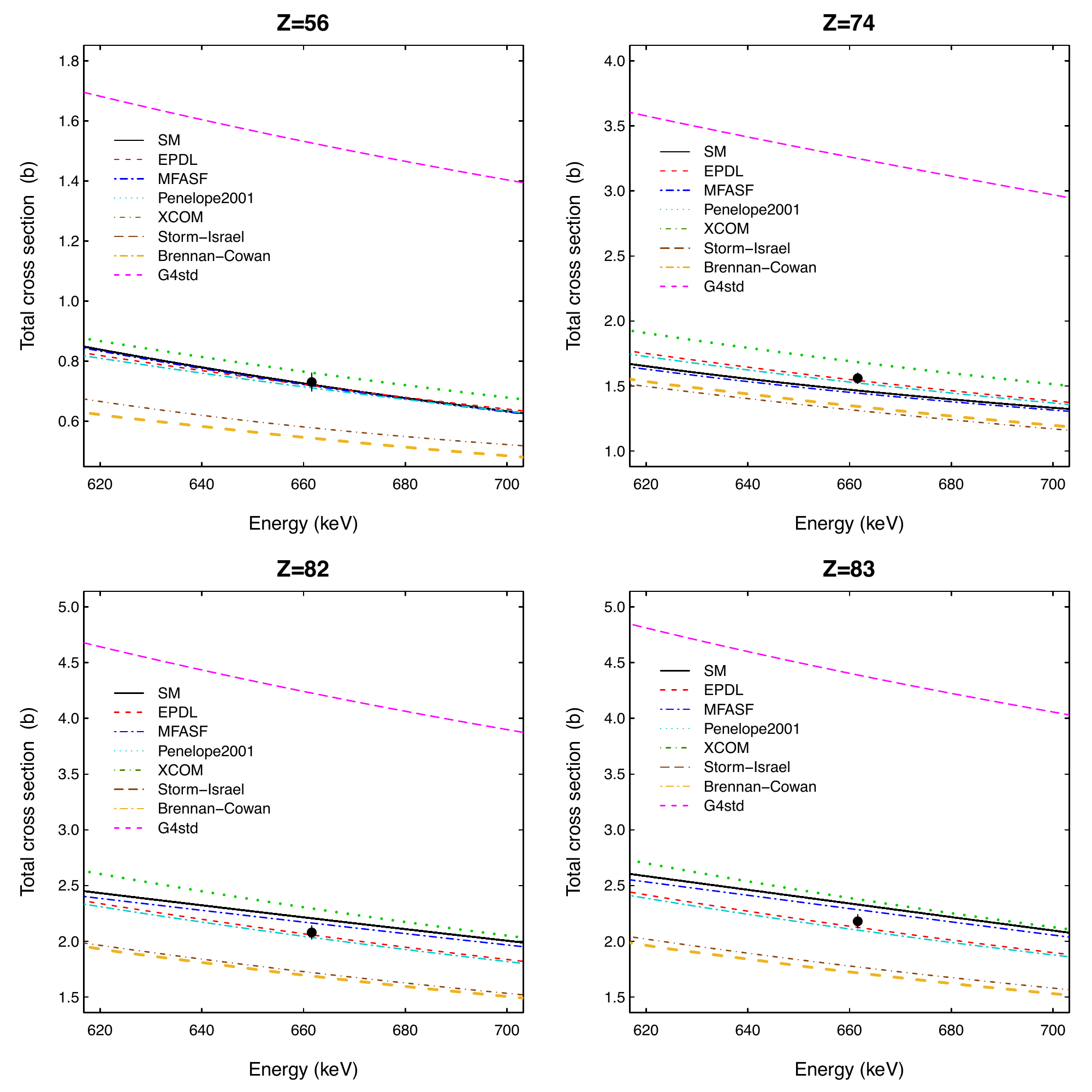}}
\caption{Total cross section as a function of energy for barium, tungsten,
lead and bismuth: calculations based on S-matrix (SM, black solid line), EPDL (red
dashed line), modified form factors with anomalous scattering factors MFASF
(blue double-dashed line), XCOM (green dotted line), Storm and Israel (brown
dotted-dashed line), Penelope 2001 (turquoise double-dashed line),
Brennan and Cowan (yellow dashed line) and 
\textit{G4XrayRayleighModel} (G4std, magenta long-dashed line), and 
experimental data (black circles) from \cite{Gowda1995}
.
In some plots the experimental errors are smaller than the symbol size.}
\label{fig_totcsexp}
\end{figure*}

Although in the higher energy end all the calculated cross sections look quite
similar on the scale of Fig. \ref{fig_totcs}, differences are visible in
Fig.~\ref{fig_totcsexp}, that highlights the behaviour of total cross section
models in a region where experimental data are available in the literature
\cite{Gowda1995}. 

Most of the models shown in Fig. \ref{fig_totcsexp} produce
qualitatively similar results in that energy range, apart from the
interpolation of Storm and Israel's cross sections, the
parameterizations by Brennan and Cowan, and \textit{G4XrayRayleighModel};
this model exhibits large discrepancies with respect to Gowda's measurements.
The total elastic scattering cross sections calculated by the S-matrix, EPDL and
MFASF models are compatible with the measurements in \cite{Gowda1995} with 0.01
significance; the cross sections based on Storm and Israel's tabulations and
calculated by XCOM lie respectively 2.1 and 1.1 standard deviations away from
the experimental values for barium, while they differ from the experimental values of
the other three target elements by 3.4 to 5.2 standard deviations.
The cross sections calculated by \textit{G4XrayRayleighModel} lie 26 to 41
standard deviations away from the experimental values.

\begin{figure*}
\centerline{\includegraphics[angle=0,width=18cm]{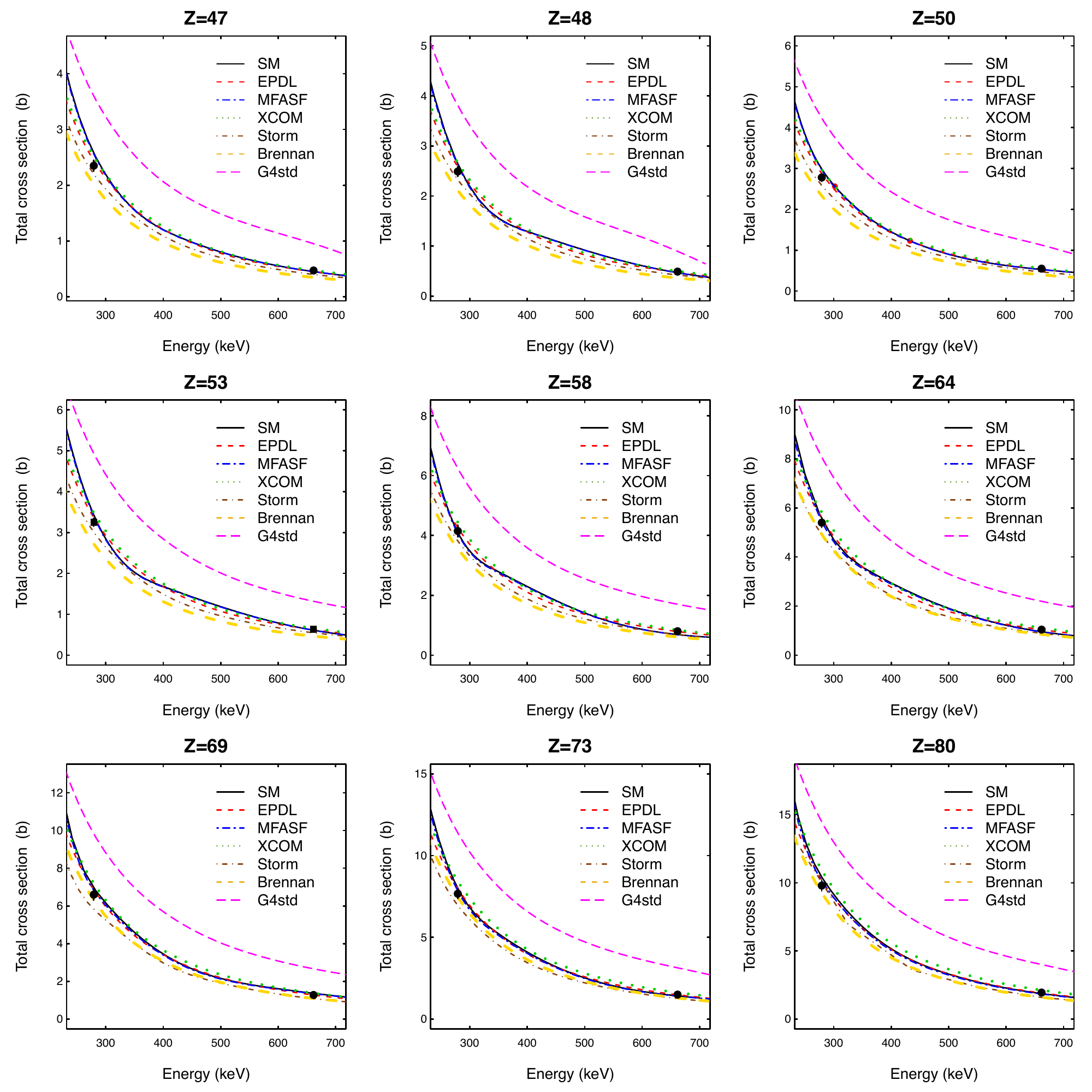}}
\caption{Total cross section as a function of energy for silver,  cadmium, tin, iodine,
cerium, gadolinium, thulium, tantalum and mercury:
 calculations based on S-matrix (SM, black solid line), EPDL (red
dashed line), modified form factors with anomalous scattering factors MFASF
(blue double-dashed line), XCOM (green dotted line) and Storm and Israel (brown
dotted-dashed line), Brennan and Cowan (yellow dashed line) and
\textit{G4XrayRayleighScatteringModel} (G4std, magenta long-dashed line) and
experimental measurements from \cite{Gowda_thesis} (black dots).
The Penelope 2001 model, not shown for better clarity of the plots, exhibits
similar behaviour to XCOM cross sections.
In some plots the experimental errors are smaller than the symbol size.}
\label{fig_totcsgowda}
\end{figure*}

Experimental data from \cite{Gowda_thesis} and total cross sections calculated
by various simulation models in the energy range between 250 and 700 keV are
in Fig. \ref{fig_totcsgowda} for 9 elements; the experimental errors of the
data at 279.2 keV reported in the figure have been scaled by a factor 10 with
respect to the values reported in \cite{Gowda_thesis}, due to the apparent 
inconsistency discussed in section \ref{sec_exp}.
The plots for the other 9 target elements measured in \cite{Gowda_thesis} 
exhibit similar behaviour to those shown in Fig.~\ref{fig_totcsgowda}.
The source of the data \cite{Gowda_thesis}, which has not been
subject to peer review for publication in a scientific journal, and the concerns
about the experimental uncertainties discussed in
section \ref{sec_exp}, suggest caution in using these data for quantitative
validation of the simulation models; nevertheless, these plots, as well as those
for the other 9 measured elements, confirm qualitatively the same conclusions
drawn from the analysis of the subset of published data.

Total cross sections based on EPDL97 appear quite similar to those deriving from
the integration of S-matrix differential cross sections, whose 
accuracy has been quantitatively assessed in section \ref{sec_diffcs}.
In the energy range between 5 keV and 1.5 MeV, which approximately corresponds
to the scope of the validation of differential cross sections documented in
section \ref{sec_diffcs}, the Kolmogorov-Smirnov test finds EPDL97 total cross
sections compatible with those based on S-matrix calculations for 72\% $\pm$ 5\%
of the elements, while the total cross sections based on modified form factors
with anomalous scattering factors are equivalent to those based on S-matrix for
97\% $\pm$ 2\% of the test elements.
The significance of these tests is 0.01.

XCOM cross sections are found compatible with those based on S-matrix in 28\%
$\pm$ 5\% of the test cases; this result can be considered representative also for a
set of models  (Storm and Israel's, Brennan and Cowan's, Penelope 2001)
that exhibit similar characteristics.


Based on these considerations, one can conclude that simulation models based on
RTAB tabulated S-matrix results, modified form factors with anomalous scattering
factors and EPDL97 are preferable for total cross section calculations to models
using Storm and Israel's and XCOM databases, or Penelope 2001 and Brennan and
Cowan's parameterizations.
It appears also justified to conclude that the cross sections calculated by
Geant4 \textit{G4XrayRayleighModel} do not represent a realistic physics model.



\subsection{Nuclear Thomson scattering}
\label{sec_nuclear}

The generation of the elastically scattered photon in general purpose Monte
Carlo codes only accounts for Rayleigh scattering; RTAB tabulations enable
accounting also for nuclear Thomson scattering in the angular distribution of
the scattered photon.
The contribution due to the nuclear Thomson amplitude is more relevant at higher
energies and large angles; as it can be observed in Fig. \ref{fig_plot_theta},
the inclusion of nuclear Thomson scattering in the simulation has significant
effects on the correct estimate of photon backscattering.

The effect of accounting for the nuclear Thomson amplitude on the accuracy of the
simulation is summarized in Table \ref{tab_nt}.
The results concern the comparison of differential cross sections with
experiment, and consist of the efficiency for two variants of the S-matrix
model, respectively accounting for Rayleigh scattering and Nuclear Thomson
scattering, or neglecting the latter; they are reported in three test
configurations, concerning the whole data sample and the data respectively below
and above 90$^{\circ}$.
In all the configurations the inclusion of nuclear Thomson scattering
contributes to improving the accuracy of the simulation.

\begin{table}[htbp]
  \centering
  \caption{Efficiencies for differential cross sections based on S-matrix 
calculations accounting for Rayleigh scattering only or including also nuclear Thomson scattering}
    \begin{tabular}{c|cc}
    \hline
          & \multicolumn{1}{c}{\textbf{SM}} & \textbf{SM  }\\
          & \textbf{Rayleigh} & \textbf{Rayleigh + Nuclear Thomson} \\
    \hline
    all   & \multicolumn{1}{c}{0.72 $\pm$ 0.05} & 0.77 $\pm$ 0.05\\
    $\theta \leq 90^{\circ}$ & \multicolumn{1}{c}{0.76 $\pm$ 0.05} & 0.82 $\pm$ 0.05 \\
    $\theta >90^{\circ}$ & \multicolumn{1}{c}{0.35 $\pm$ 0.12} & 0.59 $\pm$ 0.12\\
\hline
    \end{tabular}%
  \label{tab_nt}%
\end{table}%


\subsection{Simulation at the electronvolt scale}

Microdosimetry and nanodosimetry simulation requires the capability of modeling
particle interactions with matter down to the scale of a few electronvolts.
EPDL97 includes tabulations of Rayleigh total cross sections down to 1~eV,
nevertheless, its documentation warns about using the data below 100~eV for
photon transport calculations, since ``the uncertainty in the data rapidly
increases with decreasing energy'' \cite{epdl97}.

\begin{figure*}
\centerline{\includegraphics[angle=0,width=18cm]{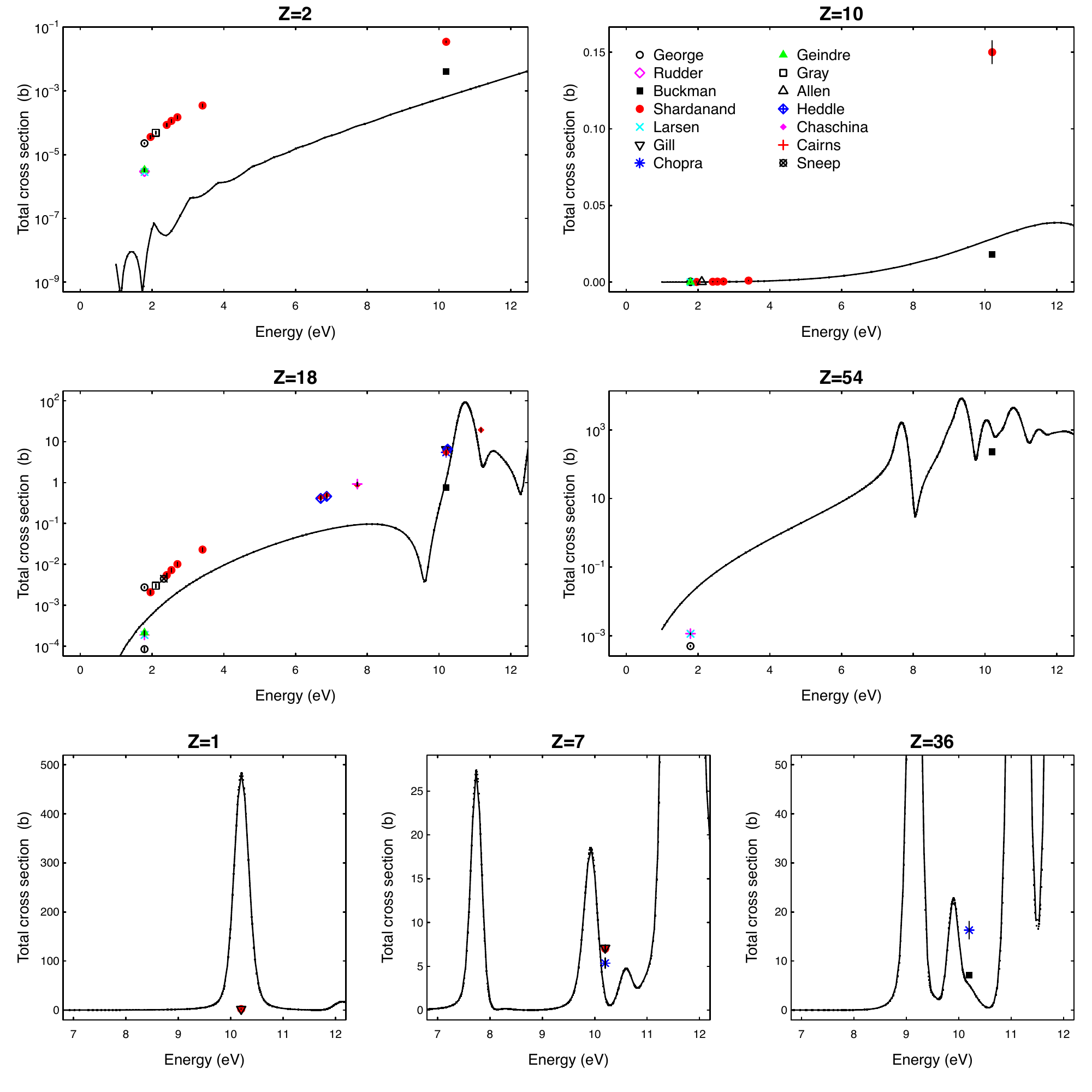}}
\caption{Total cross section as a function of energy for rare gases (helium,
neon, argon, krypton and xenon), hydrogen and nitrogen in the electronvolt
energy range: EPDL (black solid line), and experimental measurements
\cite{George1965, Rudder1968, Buckman2000, Shardanand1967, Shardanand1977, Gill1963, Chopra1974,
Geindre1973, Gray1972, Allen1973, Heddle1963, Chaschina1968, Cairns1970, Sneep2005}.
The legend reported in the plot for neon is pertinent to all the plots; the
experimental data are identified by the name of the first author of the
corresponding publication. 
Please note that the plots for helium, argon and xenon are in logarithmic scale on the
vertical axis, while the others are in linear scale.
The error bars that are not visible in the plots are smaller than the symbol size.}
\label{fig_totcsev}
\end{figure*}

A qualitative comparison of EPDL97 total Rayleigh scattering cross sections and
experimental data \cite{George1965, Rudder1968, Buckman2000, Shardanand1967,
Shardanand1977, Gill1963, Chopra1974, Geindre1973, Gray1972, Allen1973,
Heddle1963, Chaschina1968, Cairns1970, Sneep2005} for rare gases, hydrogen and
atomic nitrogen at energies between 1.8 and 11.2~ eV is illustrated in Fig.~\ref{fig_totcsev}.
Although one can observe some resemblance between EPDL97 calculations and
experimental cross section measurements in some of the plots and large
discrepancies in other ones, a quantitative appraisal of the accuracy of EPDL97
is not realistic due to the scarcity of experimental data, their inconsistencies
and possible systematic effects introduced by normalization procedures applied
to some of the experimental data.

Further experimental measurements would be needed for a thorough assessment of 
simulation models meant for microdosimetry simulation.

\subsection{Interpolation algorithms}
\label{sec_interpol}

Simulation models based on data tabulations calculate the physical quantities
needed in the course of particle transport by interpolation over the tabulated
values.
This method has the advantage of avoiding the computation of complex analytical
formulae in the course of simulation execution; nevertheless, interpolation
methods themselves could be computationally expensive, especially if they
involve calls to library functions (e.g. the calculation of logarithms) 
rather than elementary arithmetic operations only.

EPDL documentation recommends double-logarithmic interpolation over the
tabulated data; this prescription has been adopted for the related simulation
models \cite{lowe_e}.
The same interpolation method has been used also for the other models based on 
form factor tabulations to ensure a uniform treatment.

The use of S-matrix tabulations requires two-dimensional interpolation over
energy and angle for any given target atomic number. 
Three interpolation algorithms have been evaluated, considering both the
resulting physics accuracy and their computational performance: linear and
logarithmic interpolation, and a simplified interpolation method described in
\cite{rtab}, which utilizes scaling factors related to MFASF tabulations.
Table \ref{tab_interpol} summarizes the efficiency at producing results 
consistent with experiment associated with various interpolation method;
they concern the Rayleigh scattering amplitude only, but the conclusions hold 
also for other amplitudes.
The logarithmic interpolation algorithm determines differential cross sections
that are compatible with experiment at 0.01 level of significance in a larger
number of test cases; nevertheless, the efficiencies resulting from the three
algorithms are compatible within one standard deviation.

Although less accurate than logarithmic interpolation, linear interpolation of
S-matrix tabulations produces estimates in better agreement with experiment than
models based on the form factor approximation; therefore, if computational
performance is a concern, one can opt for linearly interpolating S-matrix
tabulations, still preserving the superior accuracy of this model with respect
to other physics approximations despite some degradation with respect to 
logarithmic interpolation.

Programming techniques for performance optimization
\cite{mincheol_mc2010,mincheol_nss2010,mincheol_chep2010} and
more refined interpolation methods than those discussed in this paper
can be applied for efficient tabulated data management; their in-depth
discussion is outside the scope of this paper.

\begin{table}[htbp]
  \centering
  \caption{Effect of interpolation algorithms }
    \begin{tabular}{l|cccc}
    \hline
         & {\textbf{SM}} & {\textbf{SM}} & {\textbf{SM}}  \\
                              & {\textbf{linear}} & {\textbf{logarithmic}} & {\textbf{simplified}}  \\
    \hline
     Test cases & 71    & 71    & 71     \\
          Pass      & 49    & 51    & 48     \\
          Fail      & 22    & 20    & 23    \\
          Efficiency   & 0.69  & 0.72  & 0.68   \\
          Error & $\pm$0.05   & $\pm$0.05   & $\pm$0.05   \\
    \hline

     \end{tabular}%
  \label{tab_interpol}%
\end{table}%

%


\subsection{Computational performance}

Computational performance is an important issue in assessing whether a physics
model is suitable for Monte Carlo particle transport, and in optimizing
the selection of physics models for the simulation of a given experimental
scenario, when multiple options are available.

The validation analysis documented in section \ref{sec_diffcs} has identified
the simulation model based on RTAB tabulations of second order S-matrix
calculations as the most accurate at reproducing experimental measurements
of elastic scattering differential cross sections, i.e. at determining the
actual outcome of the scattering process. 
Nevertheless, the use of S-matrix tabulations in particle transport requires
more complex operations than models based on the form factor approximation,
since it implies the management of two-dimensional tabulations (over a grid of
energies and angles) and the development of an ad hoc algorithm to sample the
direction of the scattered photon, replacing the sampling algorithms
that are currently established to deal with the form factor approximation.
The comparative evaluation of its computational performance with respect
to models based on the form factor approximation is relevant to determine
whether this modeling approach, that has not yet been exploited in Monte Carlo 
codes, is practically usable in that context.

From another standpoint, less accurate models, for instance based on
parameterizations or neglecting some physical features, could be justified for
inclusion in a simulation system, if they provide significantly faster
computational performance than more accurate alternatives.

The analysis addresses various issues related to how features specific to
different photon elastic scattering modeling approaches and software algorithms
affect the computational performance of the simulation.
Evaluations of computational performance are necessarily specific to a given
Monte Carlo code, computing platform and test case; the results reported here
concern test cases using the Geant4 toolkit, and have been obtained on an AMD
2.4 GHz 2-core processor computer with Scientific Linux 5 operating system and
the gcc 4.1.2 compiler.

From a computational perspective, one can distinguish two methods of total cross
section calculation as a function of energy: either by interpolation of
tabulated values, or by means of analytical formulae.
In the context of Geant4 the former method is used by the models based on
EPDL97, while the latter is used by \textit{G4XrayRayleighModel} and by the
model derived from Penelope 2001.
Total cross sections based on S-matrix calculations are computed by
interpolation of tabulations as well; therefore, in computational terms, they do
not differ from calculations currently based on EPDL97.

Two test cases have been evaluated as representative of the computational burden
associated with either method; they concern the average time to calculate one million total
cross sections, for atomic number between 1 and 99, and photon energy between
250 eV and 50 keV.
The computation time is $0.277 \pm 0.019$~seconds for a model based on the
interpolation of EPDL97, and $0.332 \pm 0.007$~seconds for the analytical
algorithm implemented in \textit{G4XrayRayleighModel}; it appears independent
from the photon energy.
The hypothesis of equivalent computational impact of the two calculation methods is 
rejected 99\% confidence level in this test case, i.e. the calculation of total cross sections
based on the interpolation of EPDL97 is significantly faster that the calculation
performed by the simplified \textit{G4XrayRayleighModel}.
However, this result should not be generalized to the computational equivalence
of interpolation and analytical algorithms in absolute terms, since the relative
performance of these calculation methods depends on the computational complexity
of the analytically formulae and on the characteristics of the interpolation
algorithms subject to test.

Nevertheless, the computational impact of total cross section calculation
methods on Monte Carlo simulation is marginal, since in general it affects the
computational performance only in the initialization phase of the simulation:
for instance, in Geant4 atomic total cross sections are calculated at
initialization in the process of building mean free path tabulations for each
material present in the geometrical model of the experimental set-up, which are
then used in the course of particle transport.

Two issues contribute to the computational performance of the actual photon
scattering simulation: the calculation of the physical distribution to sample
the scattering angle, and the sampling algorithm.
The calculation of the distribution to sample can occur by interpolation of
tabulated values or through an analytical formula.

The computational performance of a set of representative methods for the
generation of the elastically scattered photon has been analyzed on a test case
consisting of one million primary photons with energy between 250 eV and 1 MeV,
which are only subject to elastic scattering as possible interaction with
matter. The Geant4-based application developed for this test can be configured
to use any of the existing Rayleigh scattering options in Geant4 9.5 or the
newly developed software described in this paper, namely elastic scattering
based on S-matrix calculations; it provides options for elemental targets with
atomic number between 1 and 99.

\begin{figure*}
\centerline{\includegraphics[angle=0,width=18cm]{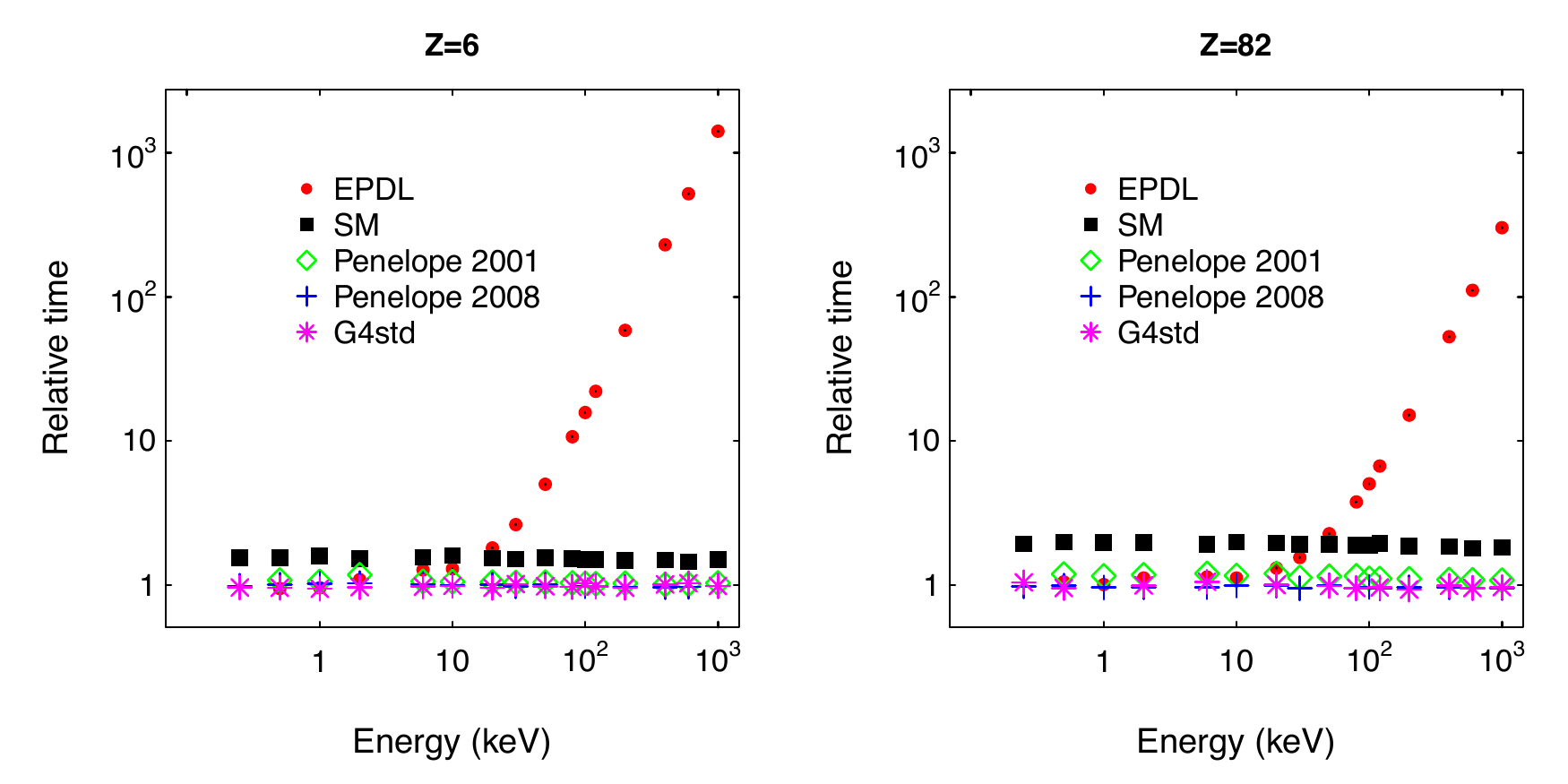}}
\caption{Time spent to simulate one million photon elastic scatterings with
carbon (left) and lead (right), as a function of photon energy with current Geant4 Rayleigh scattering
implementations and with the new model based on S-matrix calculations:
reengineered Penelope 2001 and 2008 models (respectively green diamond and blue
cross), EPDL-based "Livermore" model (red circles), \textit{G4XrayRayleighModel}
(magenta asterisk) and S-matrix model (SM, black squares).
The timing is expressed in relative units with respect to the time spent to simulate one
million scatterings with copper at 10 keV using the Geant4 model reengineered from Penelope 2008. 
The error bars are not visible in the plot, as they are smaller than the symbol size.}
\label{fig_time}
\end{figure*}

The results of this test are illustrated in Fig. \ref{fig_time}, 
which report the average time to simulate one million events as a function of
photon energy for two representative test cases (carbon and lead as targets);
the plotted values are scaled to the time needed to simulate one million events with
10 keV primary photons and copper as a target.

The computational perfomance of physics models based on the form factor
approximation appears independent from the way the sampling distribution is
calculated, but is strongly affected by the efficiency of the sampling
algorithm.
It is comparable for the Penelope 2001 and 2008 scattering
implementations, which adopt the same sampling algorithm (based on inverse
transform adaptive sampling \cite{penelope2001}), but calculate the sampling
distribution respectively from an analytical formula and by interpolation.
It exhibits large differences for the Geant4 Penelope 2008 and Livermore
implementations, which both interpolate EPDL97 tabulations, but use different
sampling algorithms (acceptance-rejection sampling for the Livermore model, a
form of inverse transform sampling for Penelope \cite{penelope2008}).
The inefficiency of the acceptance-rejection algorithm shows a strong dependence
on the photon energy, which severely affects the computational performance of the 
elastic scattering simulation above approximately a few tens keV.

The simplified physics model implemented in \textit{G4XrayRayleighModel} does not 
exhibit any significant computational advantage with respect to the interpolation of
EPDL97 with inverse transform sampling, which is available in Geant4 through the
reengineered Penelope 2008 model.

The generation of photon elastic scattering based on S-matrix calculations is
approximately a factor two slower than the method based on the form factor
approximation exploiting inverse transform sampling, which, among the test cases
illustrated in Figs. \ref{fig_time}, is represented by the
reengineered Penelope 2008 model: the difference is due to performing
interpolation over bi-dimensional tabulations rather than one-dimensional
interpolation.
The final state generation model based on S-matrix calculations exploits
the inverse transform algorithm to sample the scattering angle, similarly to the 
Penelope models; thanks to its more efficient sampling algorithm, it is
significantly faster than the model based on EPDL97 form factors currently
implemented in Geant4, which utilizes an acceptance-rejection sampling
algorithm.

A factor two degradation in the computational performance of photon elastic
scattering simulation can be considered an acceptable compromise for most
experimental applications requiring accurate simulation of photon interactions,
given the superior physical performance of this model demonstrated by the 
validation documented in section \ref{sec_diffcs}.

It should be stressed that the degradation of performance discussed in this
context is limited to photon elastic scattering, while in experimental practice
other types of photon interactions, and of their secondary products, are usually
activated in the simulation: the factor two penalty in computational performance
represents an upper limit, which corresponds to applications only involving
photon elastic scattering.
The relative impact of photon elastic scattering on the overall computational
performance cannot be quantified in general terms, since it depends
on many physical and user parameters specific to a given experimental scenario:
some factors affecting the computational speed of a simulation
are the energy of the particles involved,
the materials of the experimental set-up, the physics processes and models 
activated in the simulation, secondary production thresholds and other 
user cuts, the complexity of the geometrical configuration, and the 
characteristics of the user application software, but this enumeration is not
intended to be exhaustive.

A policy class for final state generation has been developed in the context of
this study, which can exploit any form factor tabulation and uses an inverse
transform algorithm to sample the photon scattering angle.
This class provides optimal computational performance, while it substantially
improves simulation accuracy with respect to the Penelope 2008 model, still
in the form factor approximation, by using modified form factors with angle 
independent anomalous scattering factors.




\section{Conclusion}

An extensive set of models for the simulation of photon elastic scattering has been
quantitatively evaluated regarding their accuracy at reproducing experimental
measurements and their computational performance.

The analysis has identified the simulation of photon elastic scattering based on
second order S-matrix calculations tabulated in RTAB as the state-of-the-art.
This model, that accounts for Rayleigh scattering and nuclear Thomson
scattering, represents approximately a factor two improvement in compatibility
with experiment with respect to the Rayleigh scattering models currently
available in Geant4 and in other general purpose Monte Carlo codes.
The inclusion of nuclear Thomson scattering is relevant especially at higher energies
and for backscattering.
Complementary evaluation of the computational performance has demonstrated that
photon elastic scattering simulation based on S-matrix calculations is
practically feasible in a general purpose Monte Carlo system, despite its greater
computational complexity than models based on the form factor approximation
currently in use.


The results of the analysis hint that the accuracy of simulation based on
S-matrix calculations could be further improved by optimizing the tabulation
grid of RTAB for more precise interpolation in areas that are sensitive to the
underlying atomic structure.
Nevertheless, the production of an extended data library of second order
S-matrix calculations would require significant computational investment and the
collaboration of expert theoreticians.

If one prefers to base the simulation of Rayleigh scattering on the form factor
approach to avoid more complex computations required by the S-matrix model,
modified form factors with anomalous scattering factors (MFASF) appear the
preferable choice among the various form factor options examined in this paper,
although they result in degraded simulation accuracy with respect to the
state-of-the-art model based on S-matrix calculations.
The choice of the method to sample the direction of the scattered photon is
critical for computational performance: inverse transform sampling is preferable
to acceptance-rejection.

Relativistic form factors result in worse accuracy than other form factor
options (non-relativistic and modified); therefore their use in simulation
models is not encouraged.

Anomalous scattering factors calculated in the RTAB theoretical environment
contribute to improve the accuracy of cross sections based on modified and
relativistic form factors, while the anomalous scattering factors included in EPDL97
do not improve the results based on this data library.

Total and differential cross sections calculated by interpolation of pristine
EPDL97 tabulations, and by the reengineered code and reprocessed EPDL97
tabulations originating from Penelope 2008 produce equivalent physics results.
The maintenance in Geant4 of two implementations that conform to the same
software design scheme and produce equivalent physics outcome appears redundant.

The total cross sections calculated by the Penelope 2001 model neglect the
underlying atomic structure in the low energy range, and the differential cross
sections exhibit significantly worse accuracy than other models based on the
form factor approximation: the maintenance of the Penelope 2001 model in Geant4
does not appear physically justified.

Total cross sections calculated by XCOM, Storm and Israel, and Brennan and Cowan  look
qualitatively similar; they exhibit relatively limited differences with respect to 
cross sections based on EPDL97 and on S-matrix calculations above approximately
10~keV, but they appear largely deficient at reproducing effects close to
absorption edges.

The total cross sections calculated by Geant4 \textit{G4XrayRayleighModel} are
insensitive to the atomic structure at low energies, and largely inconsistent
with experimental data in the energy range of a few hundred keV.
The scattering angle distribution produced by this model corresponds to Thomson
scattering by a point-like charge, which is an unrealistic description of
photon elastic scattering.
This model does not exhibit any computational advantage with respect to 
more accurate models currently available in Geant4.
Further maintenance of this model in Geant4 should be seriously considered.



The policy-based software design adopted in this study for the simulation of
photon elastic scattering has played a key role in ensuring versatility of modeling,
at the same time greatly facilitating the verification and validation process
due to minimization of the dependencies of basic physics code on other parts of
Geant4.
This finding is relevant in view of future design evolutions of the Geant4 toolkit
to improve its reliability and to reduce the efforts invested for quality assurance 
and maintenance.

The software for the simulation of photon elastic scattering developed for this
study will be proposed for release in a forthcoming Geant4 version, following
the publication of this paper, to improve Geant4 capabilities in this physics
domain.


\section*{Acknowledgment}
The authors are grateful to Lynn Kissel for valuable information and advice, and to 
Andrew Buckley for proofreading the manuscript and useful comments.

The CERN Library, in particular Tullio Basaglia, has provided helpful assistance
and essential reference material for this study.


\end{document}